\documentclass[aip,jcp,reprint]{revtex4-2}

\usepackage{amsmath}
\usepackage{upgreek}
\usepackage{float}
\usepackage{graphicx}
\usepackage{tabularx}
\usepackage{hyperref}

\begin{document}

\title{Predicting Oxidation and Spin States by High-Dimensional Neural Networks: Applications to Lithium Manganese Oxide Spinels}

\author{Marco Eckhoff}
\email{marco.eckhoff@chemie.uni-goettingen.de}
\affiliation{Universit\"at G\"ottingen, Institut f\"ur Physikalische Chemie, Theoretische Chemie, Tammannstra{\ss}e 6, 37077 G\"ottingen, Germany.}
\author{Knut Nikolas Lausch}
\affiliation{Universit\"at G\"ottingen, Institut f\"ur Physikalische Chemie, Theoretische Chemie, Tammannstra{\ss}e 6, 37077 G\"ottingen, Germany.}
\author{Peter E. Bl\"ochl}
\affiliation{Technische Universit\"at Clausthal, Institut f\"ur Theoretische Physik, Leibnizstra{\ss}e 10, 38678 Clausthal-Zellerfeld, Germany.}
\affiliation{Universit\"at G\"ottingen, Institut f\"ur Theoretische Physik, Friedrich-Hund-Platz 1, 37077 G\"ottingen, Germany.}
\author{J\"org Behler}
\email{joerg.behler@uni-goettingen.de}
\affiliation{Universit\"at G\"ottingen, Institut f\"ur Physikalische Chemie, Theoretische Chemie, Tammannstra{\ss}e 6, 37077 G\"ottingen, Germany.}
\affiliation{Universit\"at G\"ottingen, International Center for Advanced Studies of Energy Conversion (ICASEC), Tammannstra{\ss}e 6, 37077 G\"ottingen, Germany.}

\date{\today}

\begin{abstract}
Lithium ion batteries often contain transition metal oxides like Li$_{x}$Mn$_2$O$_4$ ($0\leq x\leq2$). Depending on the Li content different ratios of Mn$^\text{III}$ to Mn$^\text{IV}$ ions are present. In combination with electron hopping the Jahn-Teller distortions of the Mn$^\text{III}$O$_6$ octahedra can give rise to complex phenomena like structural transitions and conductance. While for small model systems oxidation and spin states can be determined using density functional theory (DFT), the investigation of dynamical phenomena by DFT is too demanding. Previously, we have shown that a high-dimensional neural network potential can extend molecular dynamics (MD) simulations of Li$_{x}$Mn$_2$O$_4$ to nanosecond time scales, but these simulations did not provide information about the electronic structure. Here we extend the use of neural networks to the prediction of atomic oxidation and spin states. The resulting high-dimensional neural network is able to predict the spins of the Mn ions with an error of only 0.03\,$\hbar$. We find that the Mn e$_\text{g}$ electrons are correctly conserved and that the number of Jahn-Teller distorted Mn$^\text{III}$O$_6$ octahedra is predicted precisely for different Li loadings. A charge ordering transition is observed between 280 and 300\,K, which matches resistivity measurements. Moreover, the activation energy of the electron hopping conduction above the phase transition is predicted to be 0.18\,eV deviating only 0.02\,eV from experiment. This work demonstrates that machine learning is able to provide an accurate representation of both, the geometric and the electronic structure dynamics of Li$_x$Mn$_2$O$_4$, on time and length scales that are not accessible by ab initio MD.
\end{abstract}

\keywords{Machine Learning, High-Dimensional Neural Networks, PBE0r Local Hybrid Density Functional, Oxidation States, Spin States, Lithium Manganese Oxide Spinel, Charge Ordering, Phase Transition, Jahn-Teller Effect, Electron Hopping}

\maketitle

\section{Introduction}

Battery performance plays a key role in the technological progress of portable electronic devices.\cite{Tarascon2001, Goodenough2010} Computer simulations have become an efficient tool for a target-oriented and economic development of battery materials with improved properties\cite{Hautier2010, Sanchez-Lengeling2018} since they allow to reach a fundamental understanding of underlying atomistic processes enabling systematic advances. However, in particular lithium ion batteries pose a challenge for atomistic simulations as they often contain complex oxides as positive electrode materials, in which the transition metal ions can have different oxidation states.\cite{Nitta2015, Eckhoff2020}

An important example is the commercially used lithium manganese oxide spinel Li$_{x}$Mn$_2$O$_4$ with $0\leq x\leq2$.\cite{Thackeray1983, Thackeray1997} Li$_{x}$Mn$_2$O$_4$ has a complex electronic structure including coexisting Jahn-Teller\cite{Jahn1937} distorted Mn$^\mathrm{III}$O$_6$ octahedra and undistorted Mn$^\mathrm{IV}$O$_6$ octahedra.\cite{Rodriguez-Carvajal1998, Massarotti1999, Piszora2004, Akimoto2004} The additional e$_\mathrm{g}$ electron of Mn$^\mathrm{III}$ (t$_\mathrm{2g}^3$e$_\mathrm{g}^1$) leads to increased Mn-O distances compared to Mn$^\mathrm{IV}$ (t$_\mathrm{2g}^3$e$_\mathrm{g}^0$), i.e., the electronic structure has a large impact on the local geometry. During charge and discharge of a battery the Li content varies changing the ratio of Mn$^\mathrm{III}$ and Mn$^\mathrm{IV}$ ions. In turn, this ratio of the oxidation states affects the charge ordering of Li$_{x}$Mn$_2$O$_4$, i.e., the distribution of Mn$^\mathrm{III}$ and Mn$^\mathrm{IV}$ ions, as well as the electrical conductivity, which is based on electron hopping of the e$_\mathrm{g}$ electrons among the different Mn species.\cite{Schuette1979, Goodenough1993, Shimakawa1997, Iguchi1998} Thermally induced fluctuations of the Jahn-Teller distortions in combination with electron hopping give rise to complex phenomena like structural transitions, charge ordering transitions, and conductance. Because electrical conduction is an indispensable property of a positive electrode material, a detailed understanding of electrical transport in Li$_x$Mn$_2$O$_4$ is of central interest.

The disorder of Mn$^\mathrm{III}$ and Mn$^\mathrm{IV}$ ions resulting from electron hopping as well as the fluctuations in the spatial orientations of the Jahn-Teller distortions\cite{Piszora2004, Eckhoff2020a} on average lead to a cubic Li$_{x}$Mn$_2$O$_4$ spinel structure (space group Fd$\overline{3}$m) for temperatures above $\sim$\,290\,K in the composition range $0\leq x\leq1$.\cite{Takahashi2003, Akimoto2004} Below about 290\,K the LiMn$_2$O$_4$ spinel is orthorhombic (space group Fddd) due to an increased order in the Mn$^\mathrm{III}$/Mn$^\mathrm{IV}$ arrangement and the spatial orientations of the Jahn-Teller distortions.\cite{Akimoto2000, Akimoto2004, Piszora2004, Ouyang2009, Eckhoff2020a} For $1<x<2$, a tetragonal crystal structure (space group I4$_1$/amd) with $x=2$ coexists with the cubic form.\cite{Mosbah1983, Ohzuku1989}

Using X-ray absorption spectroscopy\cite{deGroot2001, deGroot2005, deGroot2008} or electron energy loss spectroscopy\cite{Varela2009, Zhang2010, Schoenewald2020} the e$_\mathrm{g}$ occupancy or valency of the Mn ions can be determined. However, the limited time and space resolution does not allow for an investigation of the dynamics at the atomic scale. On the other hand, oxidation and spin states are readily available in spin-polarized density functional theory (DFT) calculations. Unfortunately, ab initio molecular dynamics simulations are computationally too demanding to investigate changes in the e$_\mathrm{g}$ electron mobility with temperature, especially during the transition from the orthorhombic to the cubic crystal structure, because simulations on nanosecond time scales would be required.\cite{Eckhoff2020a}

A solution could be offered by machine learning methods. The construction of atomistic potentials representing the potential energy surfaces of materials enabling large-scale molecular dynamics (MD) simulations has received a lot of attention in recent years.\cite{Behler2016,P5673,P5793} High-dimensional neural network potentials (HDNNP),\cite{Behler2007, Behler2011, Behler2017} which are a frequently used example for machine learning potentials applicable for various materials,\cite{Behler2008, Artrith2013, Gastegger2015, Morawietz2016, Hellstrom2016, Natarajan2016, Eckhoff2019, Eckhoff2020a} enable molecular dynamics simulations up to nanosecond time scales but they only provide energies and forces while electronic structure information like oxidation states are not directly accessible. However, this information is required for a comprehensive understanding of the full complexity of Li$_x$Mn$_2$O$_4$.

In comparison to the high interest in the construction of machine learning potentials, the prediction of further atomic properties related to the electronic structure by machine learning is not as common, although also this field is rapidly expanding. Examples are the prediction of atomic charges,\cite{P2962, P4419} electrostatic multipole moments,\cite{P2391} atomization energies,\cite{P3136} polarizabilities, frontier orbital eigenvalues, ionization potentials, electron affinities, excitation energies,\cite{P4514, Li2018} redox potentials,\cite{Janet2020} nuclear chemical shifts, atomic core level excitations,\cite{P4478} or even quantum mechanical wavefunctions.\cite{P5811} Moreover, the prediction of spin-state splittings and metal-ligand bond distances was successfully performed for transition metal complexes in their equilibrium configurations.\cite{Janet2017, Janet2017a, Janet2018, Taylor2020}

Here we introduce a high-dimensional neural network (HDNN) for the prediction of oxidation and spin states solely based on the atomic structure, which is applicable in the full configuration space of the system's electronic ground state. The HDNN is trained on atomic spins obtained from spin-polarized local hybrid PBE0r-D3\cite{Sotoudeh2017, Eckhoff2020} DFT calculations. This functional was extensively benchmarked for Li$_x$Mn$_y$O$_z$ systems  and related materials in a previous study showing that the atomic spins obtained for Li$_x$Mn$_2$O$_4$ reveal coexisting high-spin Mn$^\mathrm{III}$ and Mn$^\mathrm{IV}$ ions in agreement with experiment.\cite{Eckhoff2020} The HDNN is able to complement the energies and forces in HDNNP-driven MD simulations by electronic structure information to derive oxidation states, spin configurations, and electron hopping. Using the HDNN we can study the e$_\mathrm{g}$ electron mobility of Li$_x$Mn$_2$O$_4$ which has a major impact on the electrical conductivity.\cite{Shimakawa1997, Iguchi1998} Due to its relevance for battery applications, a detailed understanding of the underlying e$_\mathrm{g}$ electron hopping is of high interest. Furthermore, oxidation and spin states are of fundamental importance for reduction-oxidation reactions such as corrosion and electrocatalysis\cite{McCafferty2010} and also for spin crossovers, for example, in the binding process of oxygen to hemoglobin.\cite{Perutz1998}

In this study we focus on the relation between the structure of the MnO$_6$ octahedra and the spins of the Mn central ions, which we represent by a HDNN. Using the HDNNP developed in a separate study\cite{Eckhoff2020a} we can perform simulations of Li$_x$Mn$_2$O$_4$ below and above the transition temperature from the orthorhombic to the cubic crystal structure. The HDNN provides the electronic spin information, which is not considered in the HDNNP yielding just the energy and forces. Using both, the HDNNP and the HDNN, we can study the transition in detail structurally and electronically at the atomic scale. The time evolution of the Jahn-Teller distortions of the Mn$^\mathrm{III}$O$_6$ octahedra as well as the electron hopping of the e$_\mathrm{g}$ electrons are studied from femto- to nanosecond timescales. Furthermore, we investigate the charge order of Li$_x$Mn$_2$O$_4$ at different temperatures revealing its complex electronic structure and the impact of finite size effects.

We start by describing the HDNN method to predict oxidation and spin states. After providing the computational details, the reference data set of the HDNN is analyzed with respect to the structure-spin relation, and the quality of the HDNN approach based on the reference data is discussed. Subsequently, the predictions are tested for different Li loadings. The discussion of the phase transition, whose structural details were explained already in a separate work,\cite{Eckhoff2020a} is extended by new insights into the electronic information. Especially, the charge ordering transition is investigated in detail. Finally, we address the electron hopping as a function of the temperature revealing the role of the phase transition on conductivity.

\section{Methods}

The neural network topology we will use here is closely related to the high-dimensional neural network potential approach of Behler and Parrinello.\cite{Behler2007, Behler2011, Behler2017} However, instead of developing a method to calculate a global property like the total potential energy of a given atomic structure $\mathbf{R}=\{\mathbf{R}_n^m\}$, we aim to predict the atomic spin $S_n^m$ for each atom $n$ of element $m$ in the system, i.e., the goal is to obtain the functional relation $S_n^m(\mathbf{R})$, which is similar to our previous work on the construction of environment-dependent atomic charges for including long-range electrostatic energies in HDNNPs.\cite{P2962,P3132} 

Like in case of HDNNPs, we cannot simply use Cartesian coordinates as input for the neural network, because the derived spins must be invariant with respect to translation and rotation of the system as well as to the permutation of chemically equivalent atoms. Therefore, the Cartesian coordinates are transformed to atom-centered symmetry functions (ACSFs) $G$,\cite{Behler2011} which meet all these requirements and describe the local chemical environments of each atom inside a cutoff radius $R_\mathrm{c}$. This cutoff has to be chosen sufficiently large to include all neighboring atoms, which are relevant for the value of the spin of the central atom. Due to their many-body nature, the number of ACSFs per atom is independent of the actual number of atoms inside the cutoff sphere, which is required for the use as input of a neural network with a fixed architecture. 

For each atom the vector of ACSFs characterizing the geometric environment is then used as input for an atomic feed-forward neural network. The great advantage of neural networks is their ability to represent multidimensional real-valued functions with in principle arbitrary precision.\cite{Hornik1989} Therefore, they are well-suited as functional form for $S_n^m(\mathbf{G}_n^m)$. We note that the atomic spin only depends on the symmetry function vector of the corresponding atom which in turn depends on all Cartesian coordinates inside the cutoff sphere.

\begin{figure}[tb!]
\centering
\includegraphics[width=0.75\columnwidth]{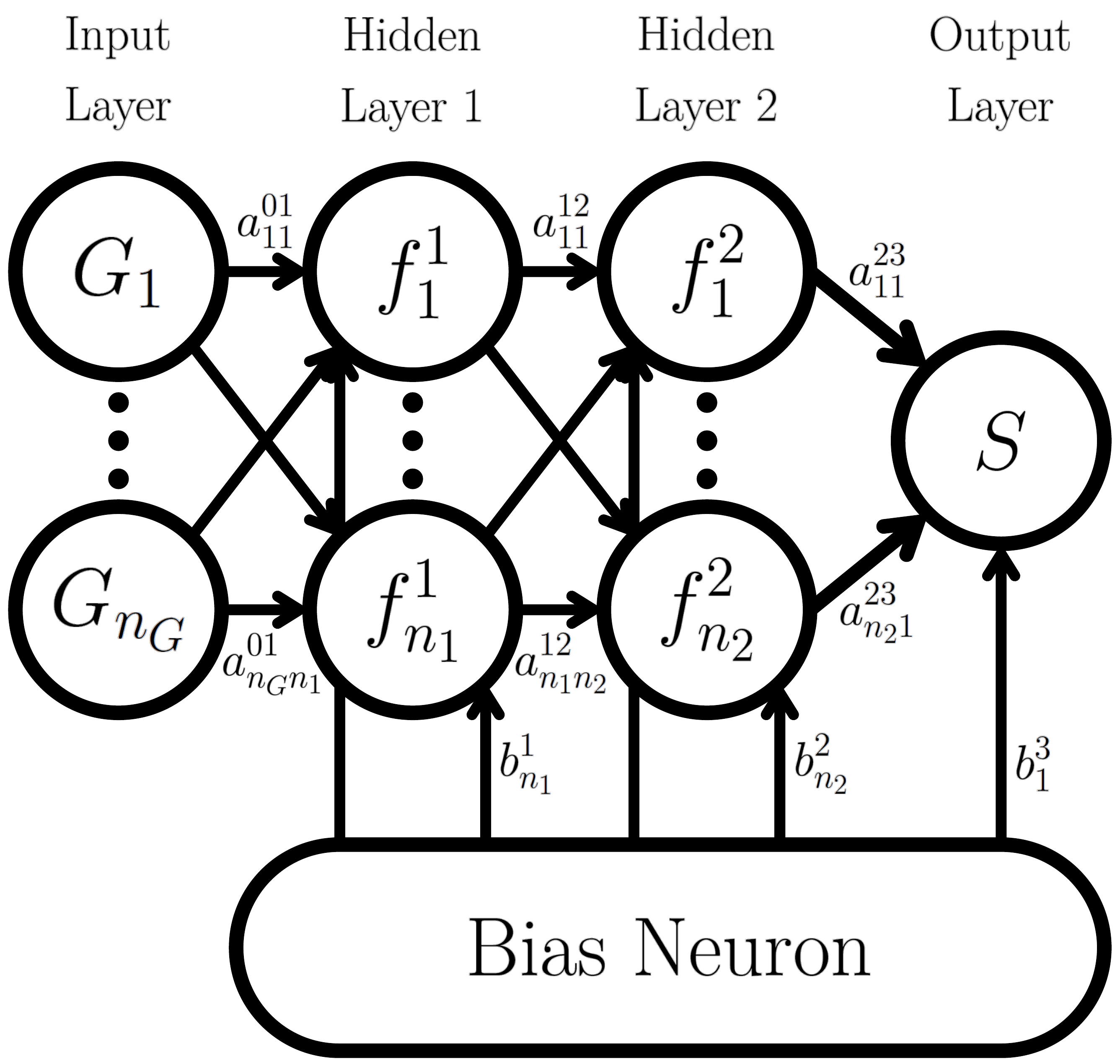}
\caption{Schematic structure of an atomic neural network containing two hidden layers that can be used to predict the atomic spin $S_n^m$ as a function of the local geometric environment described by a vector of symmetry functions $\mathbf{G}_n^m$. For clarity the subscript $n$ of the symmetry functions and atomic spin as well as the superscript $m$ of all named properties are not shown. This Figure is a visualization of Equation \ref{eq_spin}.}\label{fig_NN}
\end{figure}

The schematic structure of an atomic neural network for the calculation of atomic spins is given in Figure \ref{fig_NN}. The input layer of an atomic neural network consists of the vector of $n_G$ symmetry functions $\mathbf{G}_n^m$. This input is processed via hidden layers -- in the present case we use two hidden layers with $n_1$ and $n_2$ neurons, respectively -- resulting in a single neuron in the output layer. This output is the atomic spin of the given atom. All neurons are connected to all neurons of the neighboring layers by weight parameters $a^{\rho\sigma}_{\mu\nu}$, where $\mu$ and $\nu$ specify the two connected neurons in layers $\rho$ and $\sigma$. Additionally, all neurons in the hidden layers and the output neuron are connected to a bias neuron by a bias weight $b_{\nu}^{\sigma}$, where $\nu$ and $\sigma$ specify the number and layer of the target neuron. Together, the connecting weights and the bias weights are the fitting parameters of the neural network.

For the evaluation of each neuron in the hidden layers one proceeds from the left to the right through the network. First, the values of the neurons in the previous layer are multiplied by the corresponding connecting weights and combined, and the bias weight is added. To enable the representation of nonlinear functions, to this linear combination an activation function $f$ is applied -- in our case a hyperbolic tangent -- and the value is propagated to the next layer. For the output neuron, which finally yields the atomic spin of the given atom, the procedure is the same with the exception that the nonlinear activation function is replaced by a linear function. As a consequence, the atomic spin is given by the functional form of the feed-forward neural network,
\begin{align}
\begin{split}
S^m_n=&f_1^3\left\{b_1^3+\sum_{k=1}^{n_2}a_{k1}^{23}\cdot f_k^2\left[b_k^2+\sum_{j=1}^{n_1}a_{jk}^{12}\cdot f_j^1\Bigg(b_j^1 \right.\right.\\
&\left.\left.+\sum_{i=1}^{n_G}a_{ij}^{01}\cdot G_{n,i}^m\Bigg)\right]\right\}\hbar\ .
\end{split}\label{eq_spin}
\end{align}
The superscript $m$ is not shown for most quantities although most of them, like number of hidden layers, neurons per layer, and also the numerical values of the weights, can be different for each element $m$. 

\begin{figure}[tb!]
\centering
\includegraphics[width=0.75\columnwidth]{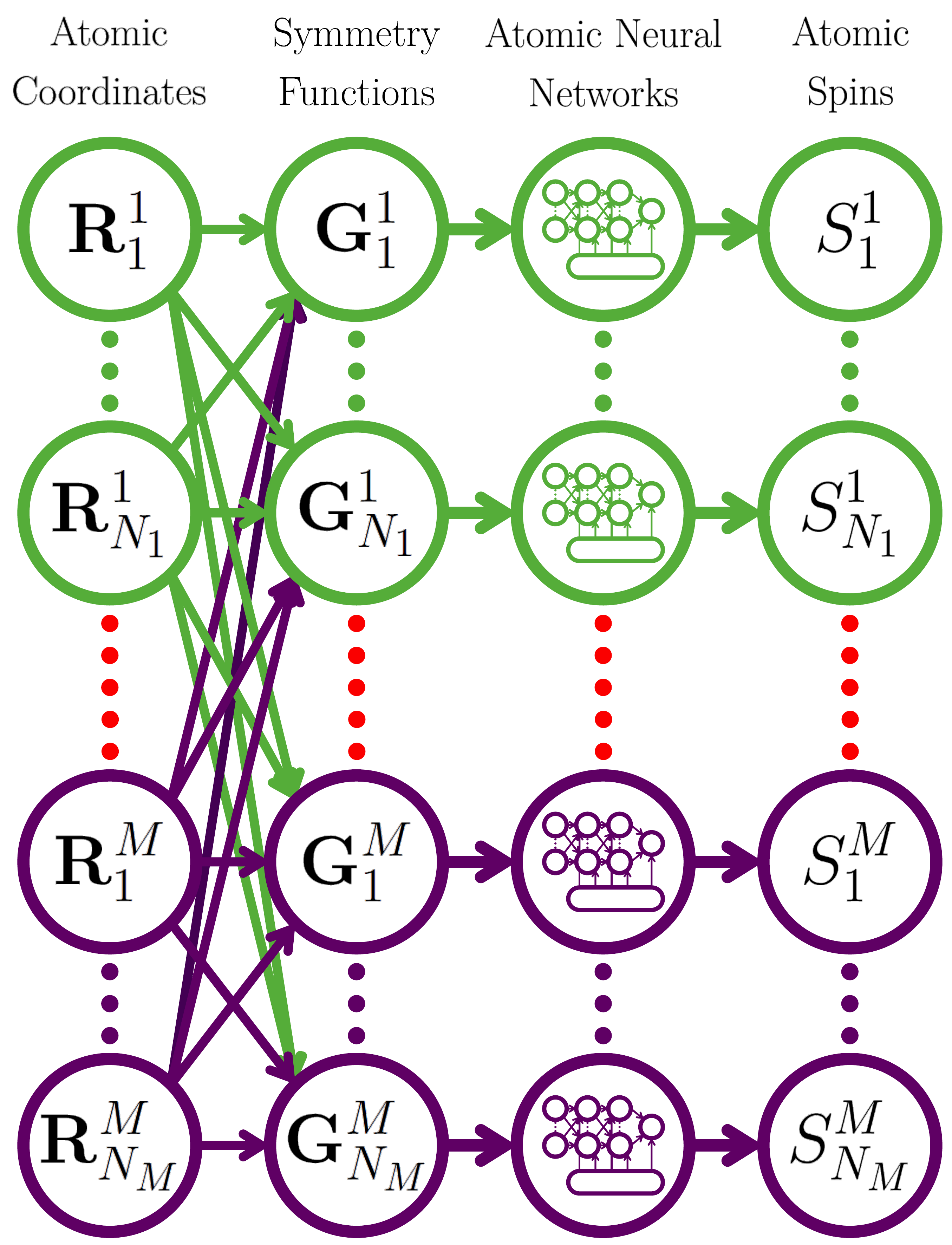} 
\caption{Structure of the HDNN for atomic spin prediction. The atomic coordinates $\{\mathbf{R}_n^m\}$ are transformed to symmetry functions $\{\mathbf{G}_n^m\}$. These are processed for each atom $n$ of element $m$ in the atomic neural network corresponding to the element yielding the atomic spin $S_n^m$ of the atom. The colors represent different elements.}\label{fig_HDNN}
\end{figure}

For each element an individual atomic neural network is constructed and replicated as many times as there are atoms of the respective element present in the system. The entire structure of the HDNN for atomic spin prediction for a system of $M$ elements and $N_m$ atoms, $1<m<M$, per element is summarized in Figure \ref{fig_HDNN}.

The crucial step is the determination of the weights $\{a^{\rho\sigma}_{\mu\nu}\}$ and $\{b_{\nu}^{\sigma}\}$ of the atomic neural networks, for which we need a reference data set containing the atomic spin values of a representative set of atomic structures. This data set can be used to iteratively optimize the weights until the neural network can reproduce the atomic spins in the configuration space spanned by the trained data with the desired accuracy. In order to check the predictive power of the HDNN and to reveal possible remaining errors, predictions for test data, which are not involved in the training process, are compared with known reference values. Only if the error of the test and the training set are comparable the predicted spins can be trusted. Additional validation steps can be applied in analogy to HDNNPs.\cite{Behler2015}

A fundamental requirement for the method to work is that the atomic structure uniquely defines the atomic spins, which is fulfilled if the correct ground state electronic structure is consistently used in all reference data. The larger the structural response with respect to changes in the atomic spins the better the method will work. Because the structure does not change if the signs of all atomic spins are inverted, we use the absolute value of the atomic spins in the training process. While in this way the information about the direction of the atomic spin is lost, all required information to determine the oxidation states and spin configurations are still available.

\section{Computational Details}\label{sec_compdetails}

The DFT reference calculations were carried out using the Car-Parrinello Projector Augmented-Wave (CP-PAW) code (version from September 28, 2016)\cite{Bloechl1994, CP-PAW} employing the local hybrid PBE0r functional.\cite{Sotoudeh2017, Eckhoff2020} All settings regarding the DFT calculations were the same as in our previous work.\cite{Eckhoff2020a} D3 dispersion corrections were applied using the DFT-D3 software (version from June 14, 2016).\cite{Grimme2010, Grimme2011} Becke-Johnson damping was used together with the settings given for the HSE06 functional according to our prior benchmark.\cite{Eckhoff2020} The atomic spins were determined by a projection of the spin density onto the one-center expansions of the partial waves. The sum of the contributions was calculated inside a cutoff radius of 1.2 times the covalent radius of the respective atom.

The HDNN for the atomic spins as well as the HDNNP for the potential energy surface of Li$_{x}$Mn$_2$O$_4$ were constructed using the RuNNer code (version from December 4, 2018 and August 22, 2019, respectively).\cite{Behler2015, Behler2017, RuNNer} The details of the HDNNP are described in our previous study.\cite{Eckhoff2020a} The atomic neural networks of the HDNN for the redox-active element Mn consists of an input layer with 63 neurons representing the ACSFs, three hidden layers with 20, 15, and 10 neurons, and an output layer with one neuron. For the elements Li and O the atomic neural networks contain 15 input neurons, three hidden layers with 4, 3, and 2 neurons, and the output neuron. The symmetry functions, which are employed for the different elements, are given in the Supplementary Material. The reference data was split into a training data set containing 90{\%} of the data and a test data set containing the remaining structures. The details of the training process\cite{Kalman1960, Blank1995} including the description of the adapted weight initialization method\cite{Glorot2010, Eckhoff2020a} are described in the Supplementary Material. The spin predictions using the final HDNN were performed by the RuNNer code.

Geometry optimizations and MD simulations driven by the HDNNP were performed using the Large-scale Atomic/Molecular Massively Parallel Simulator (LAMMPS) (version from August 7, 2019).\cite{Plimpton1995, LAMMPS} In order to use HDNNPs, LAMMPS was built with the neural network potential package (n2p2) extension (version from December 9, 2019).\cite{n2p2} The simulations were run in the isothermal-isobaric ($NpT$) ensemble applying the Nos\'{e}-Hoover\cite{Nose1984, Hoover1985} thermostat and barostat at a pressure of $p=1\,\mathrm{bar}$. The damping constants were set to 0.05\,ps and 0.5\,ps, respectively, and the simulations were done with a time step of 0.5\,fs. The angles of the simulation cell vectors were fixed at 90\,$^\circ$. The simulations were analyzed after an initial equilibration period of 1\,ns.

Basin-hopping Monte Carlo\cite{Wales1997} simulations were carried out by a self-written script using LAMMPS and n2p2 for the geometry optimizations employing the conjugate gradient algorithm. The Monte Carlo displacements were replaced by short MD runs at high temperatures in order to change the Mn$^\mathrm{III}$/Mn$^\mathrm{IV}$ distribution. This molecular dynamics based basin-hopping Monte Carlo (MDBHMC) approach enables an efficient sampling of various spatial distributions of the Mn$^\mathrm{III}$O$_6$ and Mn$^\mathrm{IV}$O$_6$ octahedra in Li$_x$Mn$_2$O$_4$.
In the MDBHMC simulations of the LiMn$_2$O$_4$ unit cell 1000 Monte Carlo steps were performed with $NpT$ MD simulation lengths of 2000 time steps each. The time step in these MD simulations was set to 1\,fs and a temperature of 400\,K has been chosen. The configurations were classified by their potential energy. If a configuration was already visited during the simulation, the MD temperature was increased by 10\,K in the next step up to a maximum value of 550\,K to escape more easily from this region of the configuration space. The Monte Carlo temperature was set to 200\,K and was increased by 100\,K if the new structure after the MD simulation was not accepted, so that the simulation does not get stuck in a low-lying minimum but still samples preferentially the configuration space at low temperatures.

\section{Results and Discussion}

\subsection{Jahn-Teller Effect}\label{sec_JT}

Our aim is to construct a HDNN, which describes the relation $S_n^m(\mathbf{R})$, i.e., the atomic spins must be uniquely defined by the geometric environments of the atoms involving the positions of many atoms. Therefore, we first have to investigate the relation between the atomic spins and the structures in the reference data set. For this purpose, we use the same reference structures that are underlying the HDNNP,\cite{Eckhoff2020a} i.e., 15228 Li$_x$Mn$_2$O$_4$ bulk structures, with $0\leq x\leq 2$. As our previous study has shown, these structures sample the configuration space of Li$_x$Mn$_2$O$_4$ accurately up to a temperature of about 500\,K. Instead of the atomic energies and forces used for the HDNNP, now the reference data set contains the absolute values of the atomic spins calculated as described in Section \ref{sec_compdetails} as the target property of the HDNN.

\begin{figure}[tb!]
\centering
\includegraphics[width=\columnwidth]{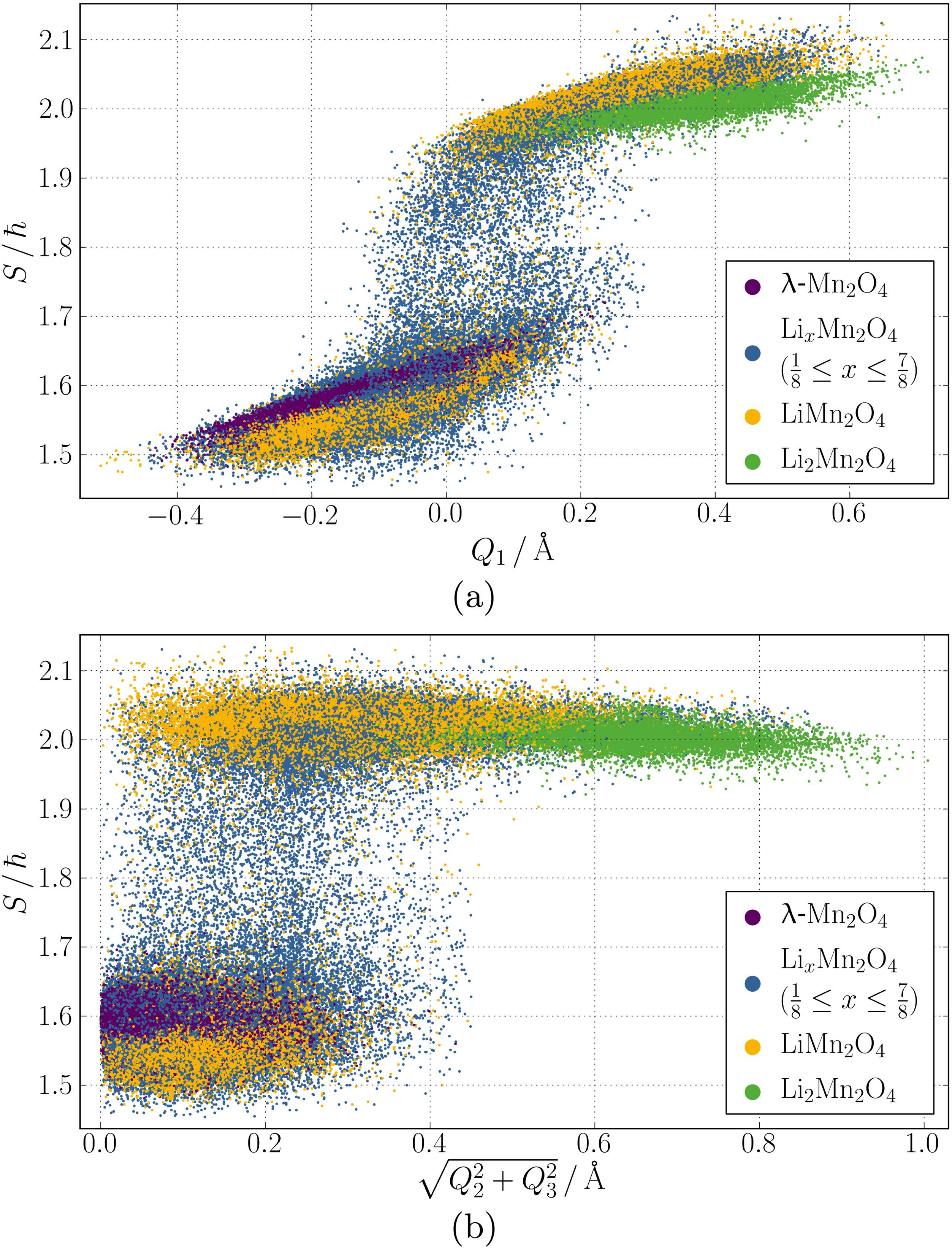}
\caption{DFT atomic spin values of the Mn ions as a function of the Jahn-Teller modes (a) $Q_1$ (Equation \ref{eq_q1}) and (b) $\sqrt{Q_2^2+Q_3^2}$ (Equation \ref{eq_g23}) of the corresponding MnO$_6$ octahedra.}\label{fig_JT}
\end{figure}

In order to show that the spins of the Mn ions in these structures are a function of the atomic environment, they are plotted in Figure \ref{fig_JT} as a function of the breathing mode and the Jahn-Teller modes of the corresponding MnO$_6$ octahedra
\begin{align}
Q_1&=\tfrac{1}{\sqrt{3}}\left(d_x+d_y+d_z-3\overline{d}\right)\ ,\label{eq_q1}\\
\sqrt{Q_2^2+Q_3^2}&=\sqrt{\tfrac{1}{2}\left(d_x-d_y\right)^2+\tfrac{1}{6}(2d_z-d_x-d_y)^2}\label{eq_g23}
\end{align}
with the distances $d_x$, $d_y$, and $d_z$ of the opposite O-O corners and the mean opposite O-O distance $\overline{d}$.\cite{Kanamori1960} Here, $Q_1$ is a measure for the size of the octahedra, while $\sqrt{Q_2^2+Q_3^2}$ refers to the Jahn-Teller distortion.

Figures \ref{fig_JT} (a) and (b) clearly show two main groups of MnO$_6$ octahedra and some intermediate data points, which is not surprising as the structures have been obtained in MD simulations containing also continuous transitions between octahedra with and without Jahn-Teller distortions due to electron hopping.\cite{Eckhoff2020a} One group is located between spin values of about 1.5 to 1.7\,$\hbar$ with a mean spin value of 1.58\,$\hbar$ corresponding to Mn$^\mathrm{IV}$ ions and the other group between about 1.9 to 2.1\,$\hbar$ with a mean of 2.01\,$\hbar$ corresponding to high-spin Mn$^\mathrm{III}$ ions. The spins values do not exactly match 1.5 and 2.0\,$\hbar$ due to the determination by a projection of the spin density inside a cutoff sphere of predefined size. This might miss contributions of the atom which are outside the cutoff or include contributions of neighboring atoms. Still, these values are qualitatively accurate enough to determine the different groups and their corresponding oxidation and spin states. Of the different structures included in the reference set, all Mn ions in $\uplambda$-Mn$_2$O$_4$ correctly belong to the Mn$^\mathrm{IV}$ group and all Mn ions in Li$_2$Mn$_2$O$_4$ to the Mn$^\mathrm{III}$ group as expected. For LiMn$_2$O$_4$, which contains Mn$^\mathrm{III}$ and Mn$^\mathrm{IV}$ in a ratio of one-to-one, both spins can be identified.

The Mn$^\mathrm{III}$O$_6$ octahedra yield larger $Q_1$ values than the Mn$^\mathrm{IV}$O$_6$ octahedra (Figure \ref{fig_JT} (a)), i.e., the Mn$^\mathrm{III}$O$_6$ octahedra are above average in size. This is in agreement with the additional occupation of the antibonding e$_\mathrm{g}$ orbital with one electron, which widens the Mn-O bonds. For about $0.0\,\mathrm{\AA}<Q_1<0.2\,\mathrm{\AA}$ both types of octahedra are present, i.e., it is not possible to reliably distinguish Mn$^\mathrm{III}$O$_6$ and Mn$^\mathrm{IV}$O$_6$ octahedra by their size. In this interval also the outliers of the two groups are found. Since they have intermediate values between the two groups, they can be classified as transition configurations.

Higher values of $\sqrt{Q_2^2+Q_3^2}$ indicate a stronger Jahn-Teller distortion of the octahedra, and for an ideal octrahedron $\sqrt{Q_2^2+Q_3^2}$ equals zero. Figure \ref{fig_JT} (b) shows that the Mn$^\mathrm{IV}$O$_6$ octahedra are less distorted than the Mn$^\mathrm{III}$O$_6$ octahedra. The reason for small distortions of the Mn$^\mathrm{IV}$O$_6$ octahedra are thermal fluctuations included in the reference data set sampled at finite temperatures. On the other hand, the Mn$^\mathrm{III}$O$_6$ octahedra show clear Jahn-Teller distortions caused by the single e$_\mathrm{g}$ electron leading to the higher values of $\sqrt{Q_2^2+Q_3^2}$. Still, like in Figure \ref{fig_JT} (a) in a certain interval the structural descriptor $\sqrt{Q_2^2+Q_3^2}$ is not sufficient for an unique identification, and the relatively low number of structures in the transition region is also hardly distinguishable by the combination of $Q_1$ and $\sqrt{Q_2^2+Q_3^2}$. 

Still, we conclude that the two simple structural descriptors $Q_1$ and $\sqrt{Q_2^2+Q_3^2}$ allow to distinguish the chemical environments of Mn$^\mathrm{III}$ and Mn$^\mathrm{IV}$ in the vast majority of structures. Therefore, in principle it should be possible to predict the spins of these species using the HDNN. The employed ACSFs are more sophisticated structural descriptors. They do not only consider the opposite corners of the MnO$_6$ octahedra, but as many-body functions they describe for any atom of interest the positions of all neighboring atoms including Li, O, and Mn within the atomic environment of radius $12\,a_0$. Therefore, they can be expected to resolve the structural differences of the octahedra in more detail than $Q_1$ and $\sqrt{Q_2^2+Q_3^2}$. Nevertheless, we expect that neither simple descriptors nor ACSFs will allow for the unique assignment of the oxidation states of all intermediate transition structures, as they represent a continuum of possible distortions and thus in principle cannot be physically assigned to one or the other species.

\subsection{High-dimensional Neural Network Spin Prediction}

\begin{figure}[tb!]
\centering
\includegraphics[width=\columnwidth]{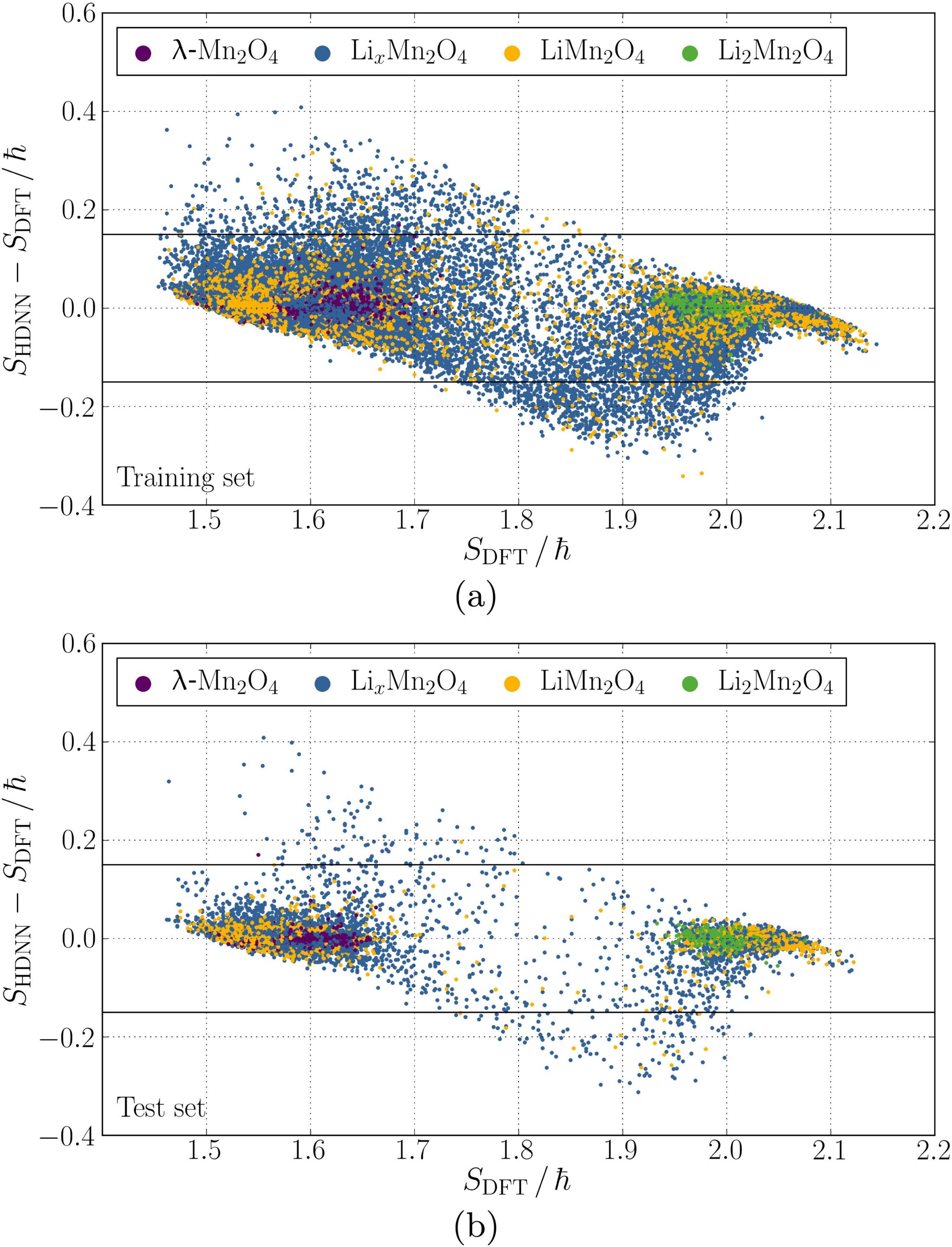}
\caption{Signed errors of the atomic spins for the (a) training set and (b) test set of the HDNN as a function of the respective DFT data. The color of the data points refers to the Li content of the structures (blue: $\tfrac{1}{8}\leq x\leq\tfrac{7}{8}$). Data points in between the two horizontal black lines have an absolute error smaller than 0.15\,$\hbar$.}
\label{fig_NN_quality}
\end{figure}

Having shown that the different spins of the Mn ions are in principle distinguishable based on the shape of the octahedra, we now aim to represent these spins by the HDNN.
Starting from the available reference set, we used the atomic spins of 13669 randomly chosen reference structures for training and the remaining 1559 structures for testing. Therefore, in total 682431 atomic environments are included in the training set and 77911 in the test set. More detailed information about the reference data can be found in the Supplementary Material of a separate work.\cite{Eckhoff2020a}

The atomic spins cover a range from 0 to 0.117\,$\hbar$ for the Li and O ions, i.e., only Li$^\mathrm{I}$ and O$^\mathrm{-II}$ ions are present, and from 1.454 to 2.144\,$\hbar$ for the Mn ions, i.e., Mn$^\mathrm{IV}$ and high-spin Mn$^\mathrm{III}$ ions. The small spins on the Li and O ions are a numerical result of the method we employ to determine the reference spins in the DFT calculations, which are not uniquely defined similar to atomic partial charges. The difference between the DFT spin density projections and the expected zero spin values, which is up to 0.1\,$\hbar$, can thus be considered as the uncertainty of our target property.  However, the spin difference of 0.5\,$\hbar$ for the two oxidation states of the Mn ions is a physically reasonable difference and suitable as target property for the HDNN.

The resulting root mean squared error (RMSE) of the spin for the Li and O ions is 0.010\,$\hbar$ for the training structures and 0.010\,$\hbar$ for the test structures indicating that the HDNN representation does not add significant errors in addition to the DFT uncertainty. For the Mn ions the corresponding values are 0.028 and 0.030\,$\hbar$ corresponding to a small relative error compared to the difference of 0.5\,$\hbar$ between the two oxidation states. The maximum spin deviation for Mn is 0.408\,$\hbar$ in the training set as well as in the test set. 0.90{\%} of Mn ions in the training data and 1.09{\%} of Mn ions in the test data have an error greater than 0.15\,$\hbar$. The deviations between the HDNN and DFT atomic spins are shown in Figure \ref{fig_NN_quality} (a) and (b) for both data sets as a function of the DFT atomic spin.

\subsection{Lithium Intercalation}

The most important process in battery applications of Li$_x$Mn$_2$O$_4$ is Li (de)intercalation, which occurs in charging and discharging. The Li content determines the ratio of Mn$^\mathrm{III}$ and Mn$^\mathrm{IV}$ ions, and the number of Mn$^\mathrm{III}$ ions must be equal to the number of Li ions to ensure overall charge neutrality. However, the neural network training process did not include any explicit information about the electrons and no electron or charge conservation condition has been imposed neither for the HDNNP nor for the present HDNN. Therefore, in case of the HDNNP all the information must be included indirectly in the potential energy surface, which in the simulations must only yield configurations with a correct Mn$^\mathrm{III}$ to Li ratio. Consequently, the correct prediction of this ratio is a stringent test for our approach.

\begin{figure}[tb!]
\centering
\includegraphics[width=\columnwidth]{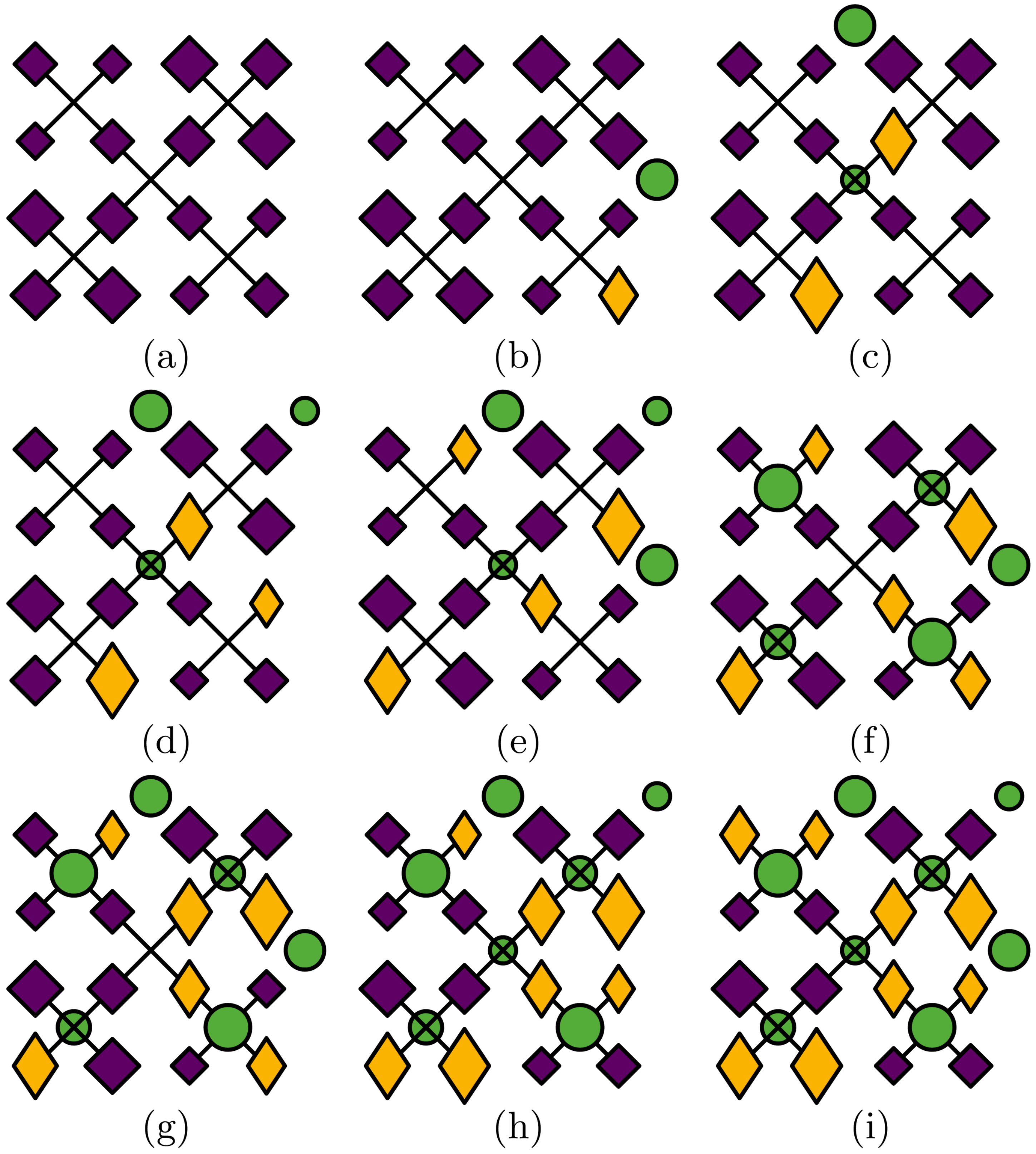}
\caption{Schematic drawing of the Li$_x$Mn$_2$O$_4$ unit cell with various Li contents (a)-(i) $0\leq x\leq1$. Li ions are shown in green, Mn$^\mathrm{III}$ ions in yellow, and Mn$^\mathrm{IV}$ ions in purple. Oxygen ions are not shown for clarity. The size of the ions represents the $z$ position from large (top layer) to small (bottom layer). The orientation of the Jahn-Teller elongated Mn$^\mathrm{III}$O$_6$ octahedra is depicted by the orientation of the yellow diamond. Jahn-Teller elongated Mn$^\mathrm{III}$O$_6$ octahedra oriented in the $z$ direction are shown as yellow squares. Black lines highlight the superstructure of corner-sharing $\left(\text{MnO}_6\right)_4$ tetrahedra.}
\label{fig_charge_order_x_111}
\end{figure}

Figures \ref{fig_charge_order_x_111} (a) to (i) show Li$_x$Mn$_2$O$_4$ unit cell structures with Li contents in the range $0\leq x\leq1$ in steps of $\tfrac{1}{8}$ optimized employing the HDNNP. The spinel structure is based on MnO$_6$ octahedra, which build a superstructure of $\left(\text{MnO}_6\right)_4$ tetrahedra. The $\left(\text{MnO}_6\right)_4$ tetrahedra share corners forming the $\uplambda$-Mn$_2$O$_4$ host lattice. For $0<x\leq1$, Li is placed in tetrahedral sites, which form a diamond structure. A three-dimensional LiMn$_2$O$_4$ representation is given in the Supplementary Material.\cite{Ovito3.2.0}

Indeed, for all Li contents the HDNN reveals that the number of Mn$^\mathrm{III}$ ions is equal to the number of Li ions. The structures with low Li content show that the Li ions are located in proximity to the Mn$^\mathrm{III}$ ions (note the $z$ direction as well as the periodic boundary conditions). In the range $0<x\leq\tfrac{1}{2}$ at most one Mn ion per $\left(\text{MnO}_6\right)_4$ tetrahedral superstructure formed by four MnO$_6$ octahedra is in the Mn$^\mathrm{III}$ state in these minimum energy configurations. At higher Li contents at least one and at most two Mn$^\mathrm{III}$ ions are found per $\left(\text{MnO}_6\right)_4$ tetrahedron.

The Jahn-Teller effect leads to elongations of all the Mn$^\mathrm{III}$O$_6$ octahedra which are aligned in the same orientation. This agrees with a previous theoretical study applying GGA$+U$ using the PBE exchange-correlation functional.\cite{Sun2020} The reported optimization of a single unit cell led as well to a tetragonal structure of LiMn$_2$O$_4$ with the conclusion that the Jahn-Teller distortions are aligned. In contrast, high-resolution diffraction experiments of the LiMn$_2$O$_4$ spinel report a very complex ordering of the Jahn-Teller distorted MnO$_6$ octahedra below the orthorhombic to cubic transition temperature.\cite{Rodriguez-Carvajal1998, Massarotti1999, Piszora2004, Akimoto2004} The orthorhombic structure exhibits five different crystallographic Mn sites. Three of these sites are predominantly attributed to Mn$^\mathrm{III}$ and two to Mn$^\mathrm{IV}$. The Jahn-Teller distortion of each Mn$^\mathrm{III}$ site is in a different direction. However, this structure does not hold the 1:1 ratio of Mn$^\mathrm{III}$ and Mn$^\mathrm{IV}$. As a consequence, eight Mn$^\mathrm{III}$ sites are occupied by Mn$^\mathrm{IV}$ ions. Representing the resulting fractional occupations in this structure would require the use of a much larger system than the $3\times3\times1$ supercell. This is not yet feasible, and our simulations employing cell sizes up to $3\times3\times3$ supercells predict an aligned state as minimum. 

\begin{figure}[tb!]
\centering
\includegraphics[width=\columnwidth]{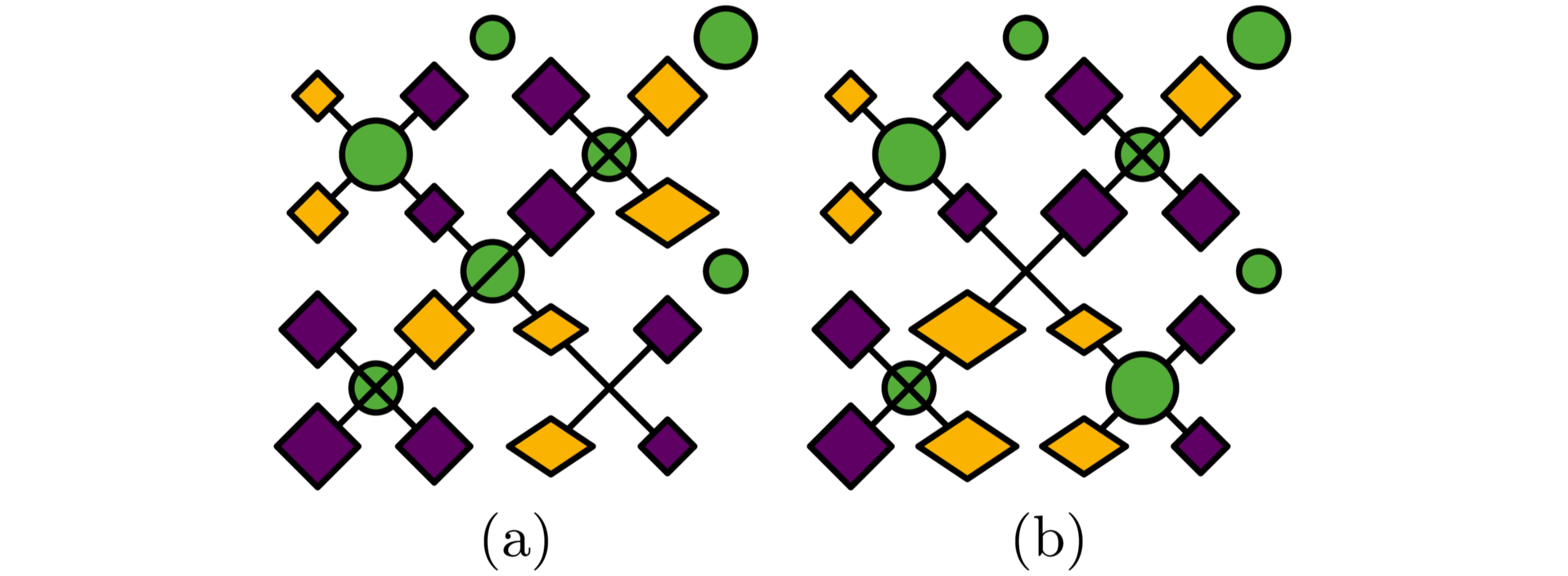}
\caption{Reorientation observed during a nudged elastic band calculation in a cubic Li$_{0.875}$Mn$_2$O$_4$ unit cell from (a) the initial minimum to (b) the final minimum. The symbols are explained in the caption of Figure \ref{fig_charge_order_x_111}.}
\label{fig_charge_order_7_111}
\end{figure}

Next, we investigate the charge reordering during lithium diffusion along a minimum energy pathway obtained from a nudged elastic band calculation\cite{Mills1994, Jonsson1998} of a cubic Li$_{0.875}$Mn$_2$O$_4$ cell in a separate work (Figure \ref{fig_charge_order_7_111}).\cite{Eckhoff2020a} While the Li ion changes from one tetrahedral site to a neighboring one, simultaneously an e$_\mathrm{g}$ electron hops from a close-by Mn$^\mathrm{III}$O$_6$ octahedron to another Mn$^\mathrm{IV}$O$_6$ octahedron. The diffusion of Li ions therefore can result in changes of the Mn$^\mathrm{III}$/Mn$^\mathrm{IV}$ distribution.

\begin{figure}[tb!]
\centering
\includegraphics[width=\columnwidth]{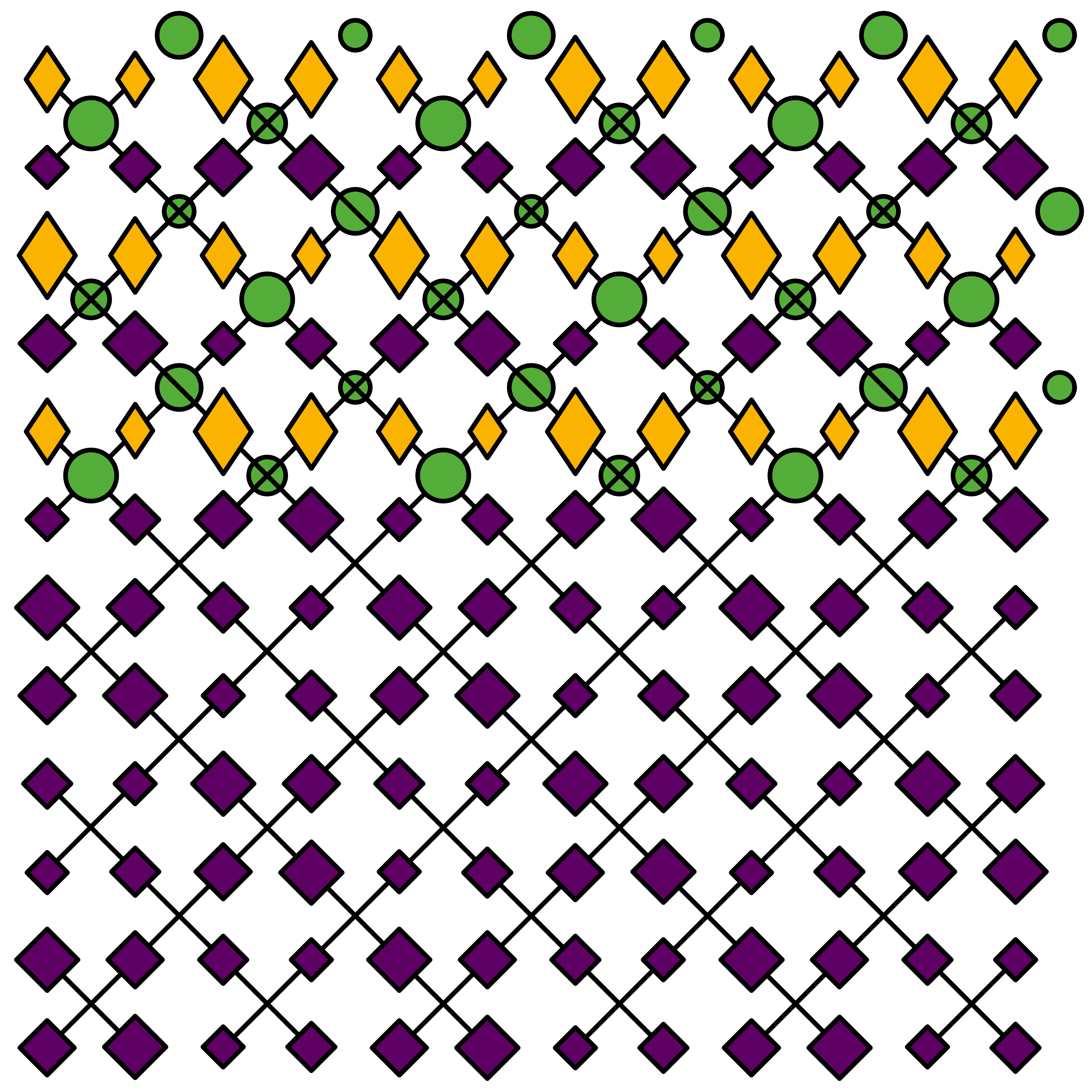}
\caption{$3\times 3\times 1$ ``layer'' of a $3\times 3\times 3$ Li$_{0.5}$Mn$_2$O$_4$ supercell, in which one half of the tetrahedral Li sites is fully occupied by Li ions, while the other half of the tetrahedral Li sites is empty. The other two ``layers'' of the $3\times 3\times 3$ supercell are identical but not shown for clarity. The symbols are explained in the caption of Figure \ref{fig_charge_order_x_111}. \label{fig_charge_order_333_1}}
\end{figure}

In order to investigate if the Li-Mn$^\mathrm{III}$ relation observed for the rather small unit cell also holds for larger systems with inhomogeneous Li distributions we constructed a $3\times3\times3$ supercell in which all tetrahedral Li sites are empty ($x=0$) for one half of the supercell and fully occupied by Li ($x=1$) for the other half. This initial structure has then been optimized to the closest local minimum employing the HDNNP, which prevents a global redistribution of the Li ions in the system. The result shown in Figure \ref{fig_charge_order_333_1} reveals that the final structure matches the expectation that Mn$^\mathrm{III}$ ions are only present in the $x=1$ half and the number of Mn$^\mathrm{III}$ is equal to the number of Li ions. We conclude that also in larger systems the Li and Mn$^\mathrm{III}$ ions are found in close vicinity.

\subsection{Orthorhombic to Cubic Transition}

Employing the HDNNP we are able to study the atomic details of the orthorhombic to cubic phase transition on nanosecond timescales. In addition, the use of the HDNN further allows to investigate the underlying electronic dynamics of the e$_\mathrm{g}$ electrons in form of the fluctuations of the atomic spins of the Mn ions. In this section, we will verify that HDNNP and HDNN yield a consistent description of the dynamics of the Mn$^\mathrm{III}$/Mn$^\mathrm{IV}$ distribution and prove the conservation of e$_\mathrm{g}$ electrons during the simulations. This will identify critical points for the prediction which are in accordance with the critical structures observed in Section \ref{sec_JT}.

The orthorhombic structure is a consequence of the Jahn-Teller distorted Mn$^\mathrm{III}$O$_6$ octahedra, which break the cubic symmetry. The transition to a cubic phase is caused by disorder and reorientations of the Mn$^\mathrm{III}$O$_6$ octahedra.\cite{Eckhoff2020a} The individual Mn$^\mathrm{III}$ octahedra are still Jahn-Teller distorted but if the dynamics are faster than the observation time or different orientations exist in the sample, the average result is a cubic structure. For unit cell simulations, no spatial averaging is given due to the restricted configuration space but due to fluctuations of the lattice constants, the time average is still cubic above the transition temperature.\cite{Eckhoff2020a}

\begin{figure}[tb!]
\centering
\includegraphics[width=\columnwidth]{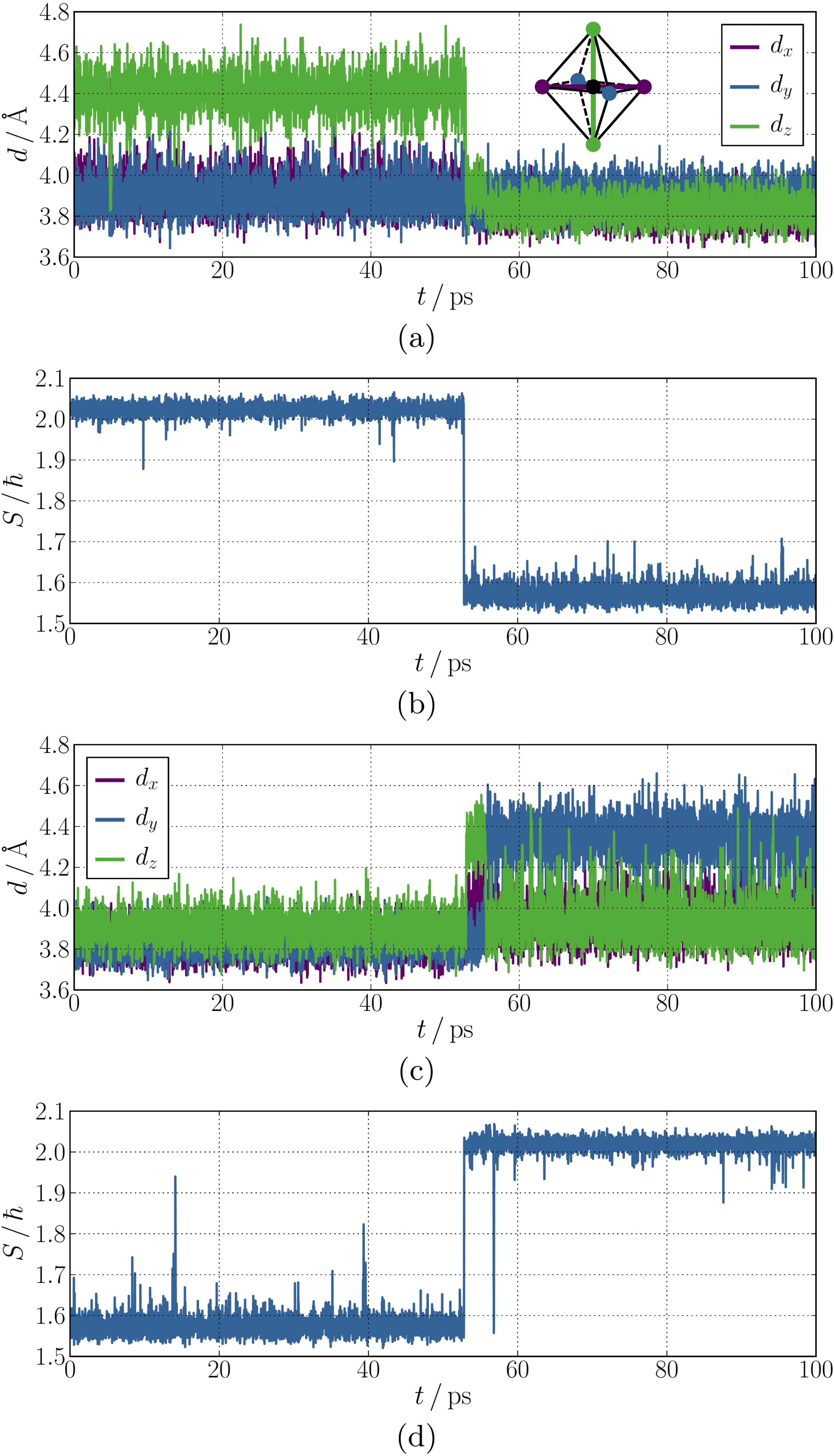}
\caption{Structural and electronic dynamics of the LiMn$_2$O$_4$ unit cell at 300\,K. Panel (a) and (c) plot the distances of the opposite corners $d_x$, $d_y$, and $d_z$ in two adjacent MnO$_6$ octahedra and (b) and (d) display the spins $S$ of the corresponding Mn ions of (a) and (c), respectively.}
\label{fig_lixmn2o4_8_111_300}
\end{figure}

The underlying atomic processes are the dynamic Jahn-Teller effect and electron hopping, which are shown in Figures \ref{fig_lixmn2o4_8_111_300} (a) and (c) by the opposite O-O distances $d_x$, $d_y$, and $d_z$ in two adjacent MnO$_6$ octahedra and Figures \ref{fig_lixmn2o4_8_111_300} (b) and (d) by the spin $S$ of the corresponding Mn ions for a simulation of the LiMn$_2$O$_4$ unit cell at 300\,K.
The e$_\mathrm{g}$ electrons hops are always to any of the adjacent Mn$^\mathrm{IV}$ sites. We observe an e$_\mathrm{g}$ electron hop from the Mn ion in Figure \ref{fig_lixmn2o4_8_111_300} (b) to the adjacent one in Figure \ref{fig_lixmn2o4_8_111_300} (d) at $t\approx53$\,ps. At the same time also the Jahn-Teller distortion of the Mn ion in Figure \ref{fig_lixmn2o4_8_111_300} (a) vanishes and a Jahn-Teller distortion of the Mn ion in Figure \ref{fig_lixmn2o4_8_111_300} (c) reemerges. The orientation of the Jahn-Teller distortion on the new site is the same as before on the old site. This is expected because the global orientation defines the preferred orientation which does not change due to a single electron hop.

At $t\approx57$\,ps, a spike is observed in Figure \ref{fig_lixmn2o4_8_111_300} (d). At first glance, this seems to be a prediction error but at the same time also the Jahn-Teller distortion of the Mn$^\mathrm{III}$O$_6$ vanishes for a short time interval of about 50\,fs (not resolved in Figure \ref{fig_lixmn2o4_8_111_300} (c)). Moreover, there are two spikes at $t\approx14$\,ps and $t\approx39$\,ps in Figure \ref{fig_lixmn2o4_8_111_300} (d). These do not reach a spin value of 2.0\,$\hbar$, i.e., they are still a transition state between Mn$^\mathrm{IV}$ and Mn$^\mathrm{III}$. In Figure \ref{fig_lixmn2o4_8_111_300} (c) the distortion of the Mn$^\mathrm{IV}$O$_6$ octahedron is also little larger than usual at these times. This is clearly no Jahn-Teller distorted Mn$^\mathrm{III}$O$_6$ octahedron but it is difficult to classify if the distortion is a result of large thermal fluctuation or if an electron transfer was about to happen.

A change of the orientation of the Jahn-Teller distorted Mn$^\mathrm{III}$O$_6$ octahedron is observed in Figure \ref{fig_lixmn2o4_8_111_300} (c) at $t\approx55$\,ps. At this time, also the global orientation of the entire system changes from $z$ to $y$ direction. In Figure \ref{fig_lixmn2o4_8_111_300} (d) no change is observable at this time, i.e., the HDNN predicts the correct oxidation state during this reorientation. Short reorientations of only the individual Mn$^\mathrm{III}$O$_6$ octahedra happen, for example, for the Mn$^\mathrm{III}$O$_6$ octahedron in Figure \ref{fig_lixmn2o4_8_111_300} (a) at $t\approx5$\,ps to the $x$ direction and for the Mn$^\mathrm{III}$O$_6$ octahedron in Figure \ref{fig_lixmn2o4_8_111_300} (c) at $t\approx62$\,ps to the $z$ direction. In both cases, the orientation changes again to the global orientation after about 0.25\,ps. The oxidation state prediction of the HDNN is in both cases correct, i.e., stays at the same spin value, during these transitions.

These results show that HDNNP and HDNN yield a consistent description of the dynamics. There might be a few prediction discrepancies for transition structures but these are in general very difficult to classify as they are in between proper Mn$^\mathrm{III}$ and Mn$^\mathrm{IV}$ states. The agreement is good even for very short processes.

\begin{figure}[tb!]
\centering
\includegraphics[width=\columnwidth]{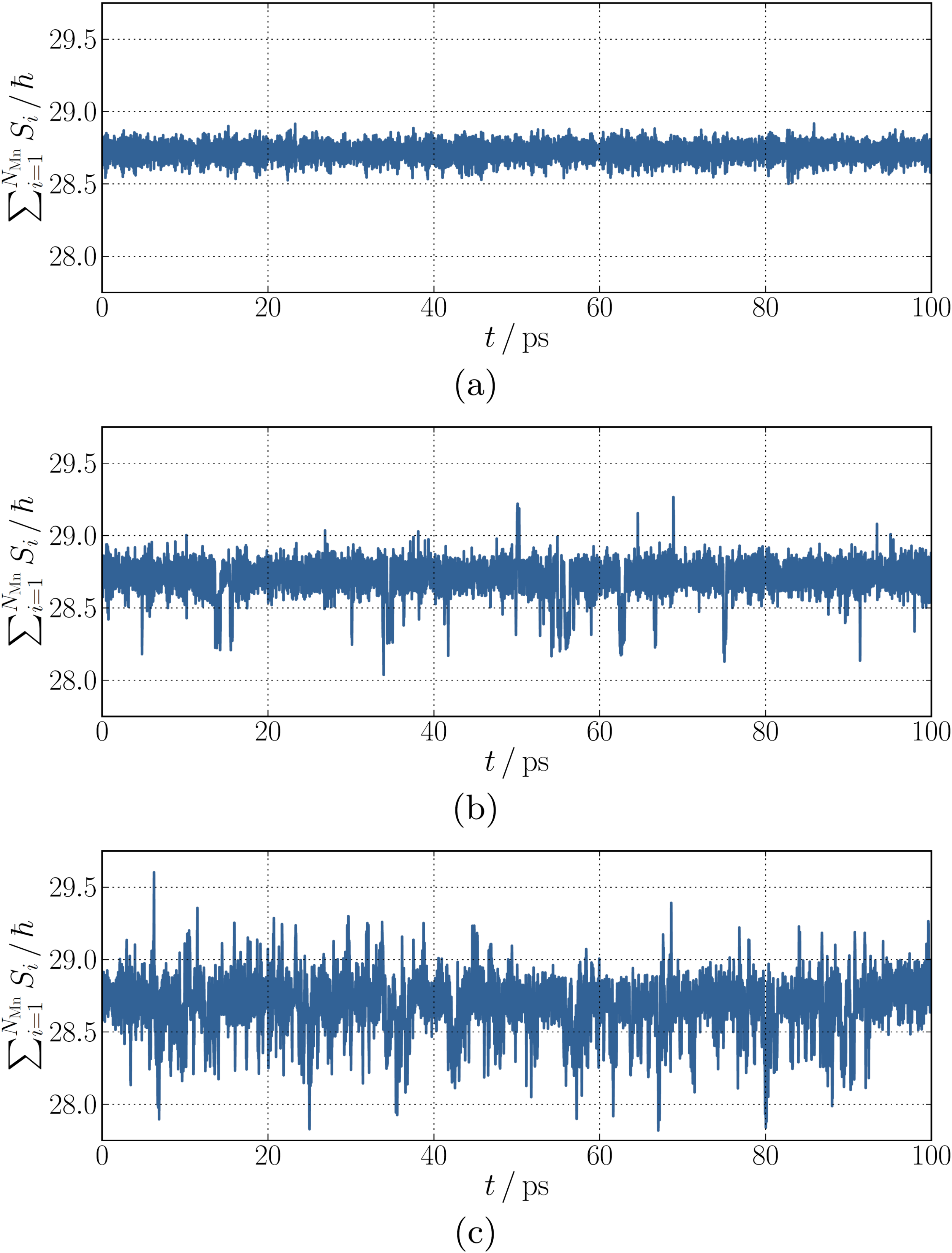}
\caption{Electronic dynamics of the LiMn$_2$O$_4$ unit cell at (a) 200, (b) 300, and (c) 400\,K represented by the sum of the Mn spins $\sum_{i=1}^{N_\mathrm{Mn}}S_i$ for the $N_\mathrm{Mn}=16$ Mn ions.}
\label{fig_lixmn2o4_8_111_S}
\end{figure}

Employing the HDNN we can also calculate the sum of the absolute spin values of all Mn ions in the unit cell at different temperatures (Figures \ref{fig_lixmn2o4_8_111_S} (a) to (c)). From ab initio molecular dynamics simulations (see Supplementary Material) we know that this property fluctuates around a mean value of about $28.8\,\hbar$ due to the conservation of e$_\mathrm{g}$ electrons. The value results from eight Mn$^\mathrm{III}$ ions with a spin value of about 1.58\,$\hbar$ and eight Mn$^\mathrm{IV}$ ions with a spin value of about 2.01\,$\hbar$, which are included in the unit cell of LiMn$_2$O$_4$. In the ab initio molecular dynamics simulations the number of electrons is constant by construction. However, there is no condition in case of the HDNN. Still, we observe in a simulation at 200\,K (Figure \ref{fig_lixmn2o4_8_111_S} (a)) that the conservation of the e$_\mathrm{g}$ electrons is excellent.

At 300 and 400\,K, where electron hopping and a dynamic Jahn-Teller effect is present, the fluctuations get a little larger but still remain close to the reasonable value of $28.8\,\hbar$. There are only a few spikes in Figure \ref{fig_lixmn2o4_8_111_S} (b) which correspond to one e$_\mathrm{g}$ electron too less or too much. One of these spikes is at $t\approx55$\,ps and a very short one at $t\approx53$\,ps. At these times the global orientation of the structure changes and the electron hop between the two Mn ions shown in Figure \ref{fig_lixmn2o4_8_111_300} happens, respectively. Reorientations of octahedra and electron hopping can cause these spikes because the transition structures include vanishing and reemerging Jahn-Teller distortions, which are very hard to classify. In combination with thermal distortions this can lead to less accurate predictions during the transition in that one Jahn-Teller distortion might already be classified as vanished but the arising one is not yet classified as a Jahn-Teller distortion or vice versa. Another source of error can be large thermal fluctuations which lead to classification uncertainties. At 400\,K (Figure \ref{fig_lixmn2o4_8_111_S} (c)) the number of these spikes increases but the dominant value is still in agreement with the expectation.

\subsection{Charge Ordering}\label{sec_charge_order}

Up to now, we used the HDNN to complement the investigation of the structural transitions by information about the atomic oxidation states and to confirm that the MD simulations correspond to a reasonable electronic structure. Now we go a step further and study electronic processes during the phase transition, which would not be possible with conventional machine learning potentials.

The charge and Jahn-Teller order are responsible for the orthorhombic crystal structure below about 290\,K. In a separate study\cite{Eckhoff2020a} we found a peak in the heat capacity at about 290\,K which we assumed to be a charge ordering transition. Employing the HDNN we can verify this assignment.

A phase transition is identified by a change of an order parameter. To describe the charge ordering transition we can employ
\begin{align}
C=\dfrac{1}{N}\sum_{t=0}^{t_N}\dfrac{\mathbf{S}_\mathrm{min}\cdot\mathbf{S}(t)}{|\mathbf{S}_\mathrm{min}|\cdot|\mathbf{S}(t)|}\ ,
\end{align}
where $\mathbf{S}_\mathrm{min}$ and $\mathbf{S}(t)$ are vectors of all Mn spins in the simulation cell in the minimum configuration and at time $t$, respectively. Both $\mathbf{S}_\mathrm{min}$ and $\mathbf{S}(t)$ correspond to the difference between the actual spin values and the mean spin value of all Mn at all times, which is about 1.8\,$\hbar$. For normalization we divide by the length of both vectors. The sum includes all time steps from $t=0$ to $t_N$ at step $N$. For $C=1$ the minimum charge order does not change with time, for $C=0$ the charge order is not correlated to the minimum charge order, i.e., the charge order fluctuates fast. The transition from $C=1$ to 0 as a function of the temperature therefore corresponds to the charge ordering phase transition.

\begin{figure}[tb!]
\centering
\includegraphics[width=\columnwidth]{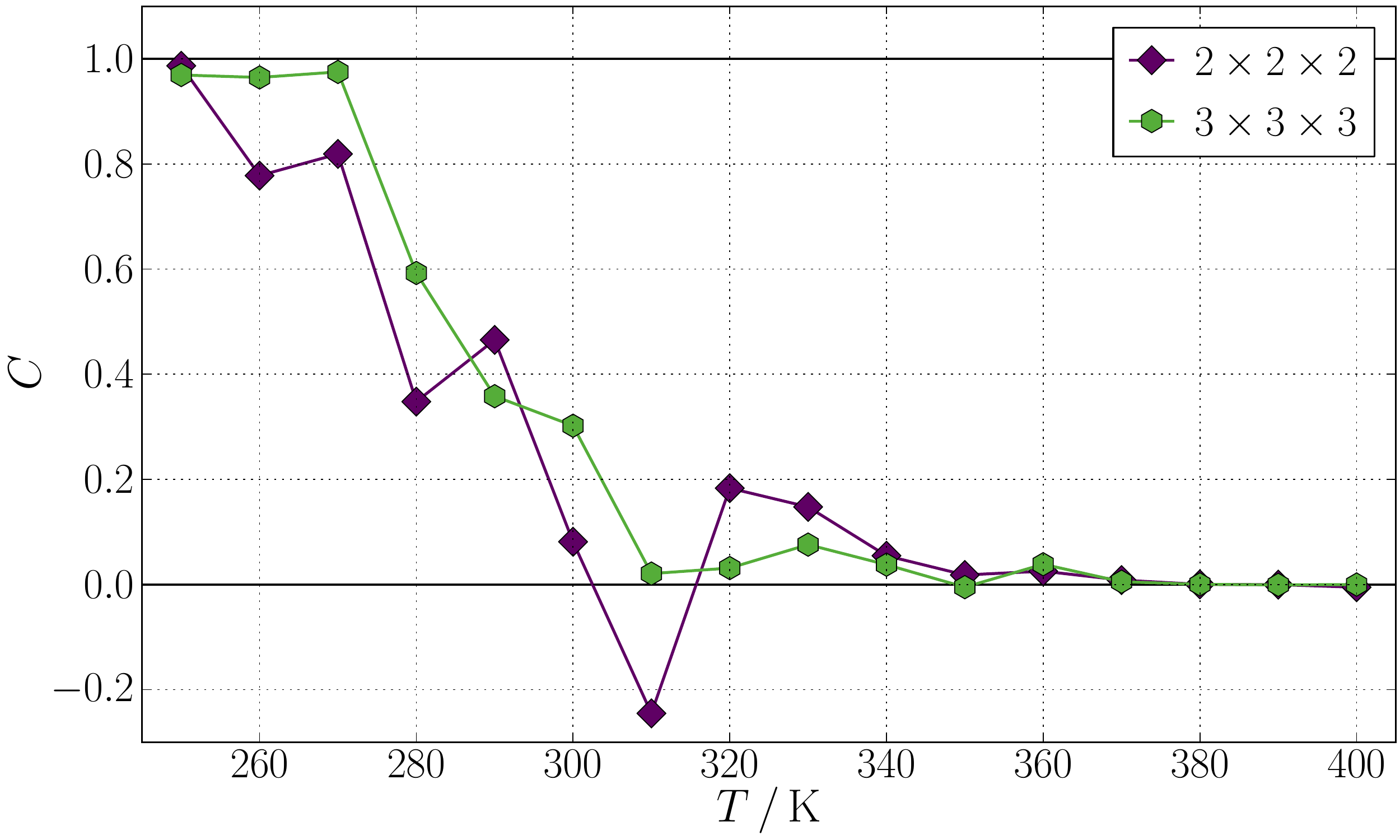}
\caption{Charge order parameter $C$ obtained from 20\,ns $NpT$ MD simulations at various temperatures of LiMn$_2$O$_4$ using different cell sizes. The black lines at $C=0$ and $C=1$ represent the ideal charge ordered and disordered phases.}\label{fig_order_parameter}
\end{figure}

Figure \ref{fig_order_parameter} shows the order parameter $C$ as a function of the temperature $T$ for the $2\times2\times2$ and $3\times3\times3$ supercell of LiMn$_2$O$_4$. The data were obtained from 20\,ns $NpT$ MD simulations. The results of the $3\times3\times3$ supercell show a clear transition between roughly 280 and 300\,K. This is in excellent agreement with resistivity experiments which yield a transition temperature of about 290\,K.\cite{Shimakawa1997} For the $2\times2\times2$ supercell the transition region is between about 260 and 340\,K. This continuous change of the order parameter over a wider temperature range is expected because of finite size effects.\cite{Challa1986, Binder1987} The transition temperature is shifted to higher temperatures with decreasing system size. In conclusion, our assumption that the peak in the heat capacity at about 290\,K is associated to a charge ordering transition is verified by the HDNN analysis.

The Mn$^\mathrm{III}$/Mn$^\mathrm{IV}$ distribution as well as the orientations of the Jahn-Teller distortions are disordered in the high temperature cubic structure. However, one still can find some frequently occurring patterns. These can also be found in simulations of a single unit cell, but of course in this case the order is the same for the entire structure. As the charge order of the low temperature phase in our simulations can also be constructed using the single unit cell we will concentrate here on the low energy charge orders of the unit cell.

\begin{figure}[tb!]
\centering
\includegraphics[width=\columnwidth]{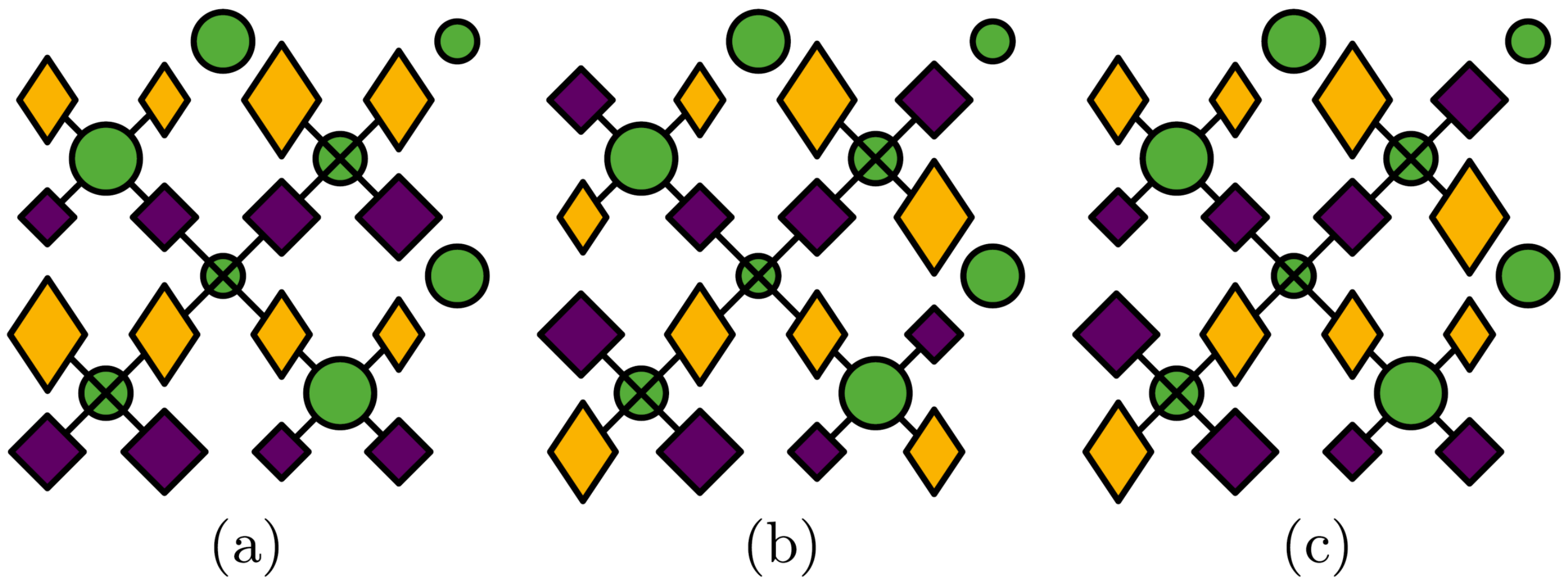}
\caption{Minimum energy configurations and orientations of the Mn$^\mathrm{III}$ ions in the LiMn$_2$O$_4$ unit cell. The potential energy increases in the order from (a) to (c). The symbols are explained in the caption of Figure \ref{fig_charge_order_x_111}.}
\label{fig_charge_order_8_111}
\end{figure}

To investigate the energetically most stable Mn$^\mathrm{III}$/Mn$^\mathrm{IV}$ distributions in more detail MDBHMC simulations were performed for a single unit cell of LiMn$_2$O$_4$. Figures \ref{fig_charge_order_8_111} (a) to (c) shows the three energetically most stable structures we found. The corresponding three charge orders agree with those found in $NpT$ MD simulations from our separate work.\cite{Eckhoff2020a} The low temperature charge order of the $2\times2\times2$ and $3\times3\times3$ supercells are based on the order in Figure \ref{fig_charge_order_8_111} (a). The energy difference between Figure \ref{fig_charge_order_8_111} (a) with lattice constants 8.664, 8.066, and 8.066\,$\mathrm{\AA}$ and Figure \ref{fig_charge_order_8_111} (b) with lattice constants 8.631, 8.113, and 8.067\,$\mathrm{\AA}$ is 1.4\,meV\,atom$^{-1}$. The difference from Figure \ref{fig_charge_order_8_111} (a) to Figure \ref{fig_charge_order_8_111} (c) with lattice constants 8.630, 8.089, and 8.089\,$\mathrm{\AA}$ is 1.9\,meV\,atom$^{-1}$. These tiny energy differences are smaller than the RMSE of the HDNNP for the test set, which is 2.2\,meV\,atom$^{-1}$, and also below the accuracy of the underlying DFT calculations. Therefore, the different charge orders are effectively degenerate in energy. In PBE0r DFT calculations for a cubic LiMn$_2$O$_4$ cell the charge order of Figure \ref{fig_charge_order_8_111} (b) is observed as global minimum.\cite{Eckhoff2020}

The distribution of Mn$^\mathrm{III}$ and Mn$^\mathrm{IV}$ ions as well as the orientations of the Jahn-Teller distortions can change during a MD simulation. Therefore, the different minimum charge orders can be found oriented in different spatial directions. The charge order of Figure \ref{fig_charge_order_8_111} (b) is the same as in Figure \ref{fig_charge_order_x_111} (i), which is not immediately obvious. In Figure \ref{fig_charge_order_x_111} (i) the Mn ions are ordered in an alternating pattern of helically aligned Mn$^\mathrm{III}$ and Mn$^\mathrm{IV}$ ions with the Jahn-Teller elongation in a direction perpendicular to the helix. A helix is formed, for example, by the four Mn$^\mathrm{III}$ ions on the right half of the LiMn$_2$O$_4$ unit cell in Figure \ref{fig_charge_order_x_111} (i). The helix includes four Mn$^\mathrm{III}$ ions per turn, i.e., the pitch is equal to the lattice constant. As the helix is aligned in $z$ direction the size of the diamonds in Figure \ref{fig_charge_order_x_111} (i) increases from the lower right Mn$^\mathrm{III}$ ion to the upper right Mn$^\mathrm{III}$ ion clock-wise with the turn of the helix. In Figure \ref{fig_charge_order_8_111} (b) the helices of Mn$^\mathrm{III}$ and Mn$^\mathrm{IV}$ ions are aligned in $x$ direction, for example, the lower four Mn$^\mathrm{III}$ ions form a helix at half height in $z$ direction.

In Figure \ref{fig_charge_order_8_111} (a) the Jahn-Teller distorted octahedra are aligned perpendicular to Mn$^\mathrm{III}$ planes resulting in a tetragonal cell. For a change of the global orientation of the Jahn-Teller distortions the Mn$^\mathrm{III}$/Mn$^\mathrm{IV}$ order has to change as already observed in a separate study.\cite{Eckhoff2020a} For the charge order in Figure \ref{fig_charge_order_8_111} (b) two degenerate configurations of the global orientation of the Jahn-Teller distortions exist. If the Mn$^\mathrm{III}$ helices are aligned, for example, in $z$ direction, the Jahn-Teller distortions can align in $x$ or $y$ direction. Moreover, also the parallel alignment to the Mn$^\mathrm{III}$ helices was observed. Therefore, the global orientation of the Jahn-Teller distortions can change without a redistribution of the Mn$^\mathrm{III}$/Mn$^\mathrm{IV}$ order. This explains the atomistic dynamics in MD simulations of a single unit cell reported in a separate work.\cite{Eckhoff2020a}

The charge order in Figure \ref{fig_charge_order_8_111} (c) seems to be an intermediate state between the other two charge orders. With only two conversions between Mn$^\mathrm{III}$ and Mn$^\mathrm{IV}$ ions each, i.e.\ two e$_\mathrm{g}$ electron hops, this charge order can transform to any of the other two charge orders. The quite small energy differences among these configurations indicate that modifications in the distributions of the Mn$^\mathrm{III}$ and Mn$^\mathrm{IV}$ ions result only in moderate energetic changes.

\subsection{Electrical Conductivity}

Electrical transport properties are of central interest for every battery material. Employing the HDNN we can study the underlying electronic processes at finite temperatures in realistic model systems which are capable to represent, for example, the charge ordering transition in LiMn$_2$O$_4$. 

The electrical conductivity,
\begin{align}\label{eq_sigma}
\sigma=e\mu n\ ,
\end{align}
is the product of the electron charge $e$ and the mobility $\mu$ and number density $n$ of the electrons which are responsible for the charge transport.\cite{Ashcroft1976} The mobility is proportional to the diffusion coefficient $D$,
\begin{align}
\mu=\dfrac{eD}{k_\mathrm{B}T}\ ,
\end{align}
with the Boltzmann constant $k_\mathrm{B}$.\cite{Bisquert2008} The diffusion coefficient in turn is proportional to the electron hopping frequency $\nu$,
\begin{align}\label{eq_D}
D=\tfrac{1}{6}d^2\nu\ ,
\end{align}
because the electron hopping occurs always to adjacent Mn$^\mathrm{IV}$ sites with the nearest-neighbor Mn-Mn distance $d$.\cite{Bisquert2004} This equation is derived for the limit of low concentrations of diffusing particles. In our case the e$_\mathrm{g}$ electrons can only hop to adjacent Mn$^\mathrm{IV}$ sites, which are only half of the Mn sites. It should be noted that the correlation between the e$_\mathrm{g}$ electrons can influence the diffusion, which is not included in this model.

Because electron hopping is a thermally activated process, the hopping frequency $\nu$ is given by
\begin{align}\label{eq_nu}
\nu=\nu_0\cdot\exp\left(-\dfrac{E_\mathrm{a}}{k_\mathrm{B}T}\right)\ ,
\end{align}
with the attempt frequency $\nu_0$ and the hopping activation energy $E_\mathrm{a}$.\cite{Mehrer2007} The activation energy for the e$_\mathrm{g}$ electron migration is a consequence of the associated Jahn-Teller distortion of the Mn$^\mathrm{III}$O$_6$ octahedra locally trapping the electrons.\cite{Schuette1979, Goodenough1993} For an e$_\mathrm{g}$ electron hop thermal distortions have to reduce the Jahn-Teller distortion of a Mn$^\mathrm{III}$O$_6$ octahedron and create a Jahn-Teller like distortion on an adjacent Mn$^\mathrm{IV}$O$_6$ octahedron. This enables the e$_\mathrm{g}$ electron to tunnel via the O ion from one Mn site to another. Subsequently, the Jahn-Teller distortion on the former MnO$_6$ octahedron vanishes and emerges on the latter one.

Combining Equations \ref{eq_sigma} to \ref{eq_nu} we obtain
\begin{align}\label{eq_Arrhenius}
\sigma T=A\cdot\exp\left(-\dfrac{E_\mathrm{a}}{k_\mathrm{B}T}\right)\ ,
\end{align}
with the pre-exponential factor,
\begin{align}\label{eq_A}
A=\dfrac{e^2d^2\nu_0n}{6k_\mathrm{B}}\ .
\end{align}

Experiments show that two thermally activated processes are relevant for the electrical conductivity of LiMn$_2$O$_4$,\cite{Shimakawa1997, Iguchi1998} which can both be described by the Arrhenius relation in Equation \ref{eq_Arrhenius}. Both processes are governed by hopping of small polarons, i.e., the e$_\mathrm{g}$ electrons of the Mn$^\mathrm{III}$ ions. The activation energy of the thermally activated hopping conduction was measured to be 0.16\,eV above the phase transition temperature and it is almost the same below the transition temperature.\cite{Shimakawa1997, Rodriguez-Carvajal1998}

To predict the electrical conductivity using the HDNN we have to investigate the dominant conduction process, which is the e$_\mathrm{g}$ electron hopping. The hopping frequency of the e$_\mathrm{g}$ electron self-diffusion can be analyzed in $NpT$ MD simulations of LiMn$_2$O$_4$ at various temperatures to obtain the $\nu(T)$ relation (Equation \ref{eq_nu}).

\begin{figure}[tb!]
\centering
\includegraphics[width=\columnwidth]{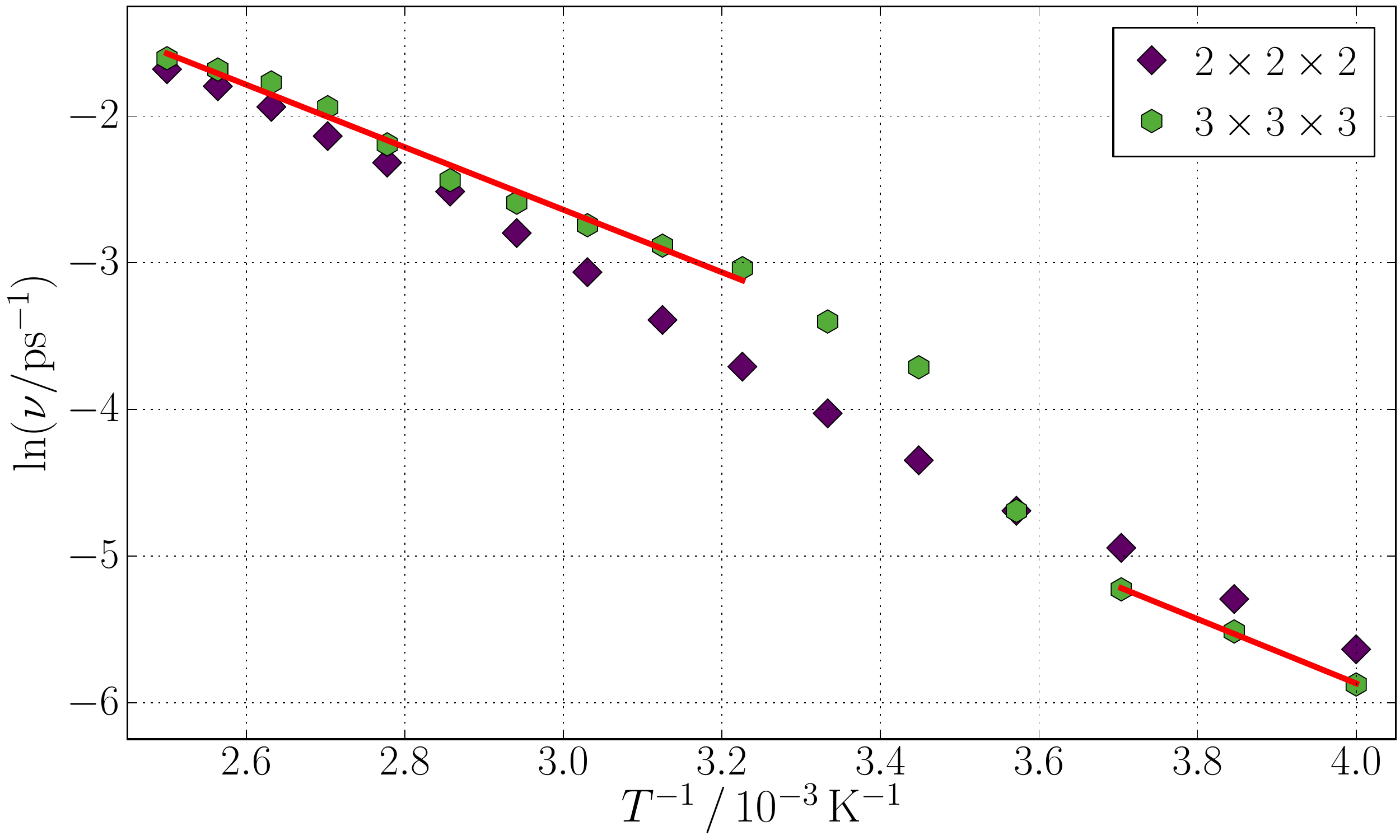}
\caption{Arrhenius relation of the $e_\mathrm{g}$ electron hopping frequency $\nu$ obtained from 20\,ns $NpT$ MD simulations at various temperatures of LiMn$_2$O$_4$ using different cell sizes. The solid lines show linear fits in different temperature regions of the $3\times3\times3$ supercell data.}\label{fig_hopping}
\end{figure}

In Figure \ref{fig_hopping} the natural logarithm of the $e_\mathrm{g}$ electron hopping frequency $\nu$ due to self-diffusion is plotted as a function of the inverse temperature $T^{-1}$. Electron hops are counted if the spin value of a Mn$^\mathrm{III}$ ion decreases below 1.65\,$\hbar$ or the spin value of a Mn$^\mathrm{IV}$ ion increases above 1.95\,$\hbar$ to exclude counting of attempted incomplete transitions which result in small spikes in $S(t)$. The absolute number of electron hops has been divided by the number of e$_\mathrm{g}$ electrons and by the total simulation time to obtain the hopping frequency $\nu$. The 20\,ns MD simulations were analyzed after each picosecond.

The results obtained for the $3\times3\times3$ supercell clearly show two different activated processes with a transition between 280 and 290\,K ($3.45\lesssim T^{-1}/10^{-3}\,\mathrm{K}^{-1}\lesssim3.57$). This agrees very well with our investigation in Section \ref{sec_charge_order} and the experimentally observed transition at about 290\,K.\cite{Shimakawa1997} The activation energy of the low temperature process (250 to 270\,K according to the order parameter in Figure \ref{fig_order_parameter}) is 0.19\,eV, while it is 0.18\,eV for the high temperature process (310 to 400\,K). This means that the transformation from the charge ordered to the charge disordered phase reduces slightly the effective energy barrier for e$_\mathrm{g}$ electron diffusion among the Mn sites.

The resulting value of $\nu_0$ obtained from the 20\,ns $3\times3\times3$ supercell data of the hopping frequency $\nu$ for the high temperature process is $\nu_0=4\cdot10^{13}\,\mathrm{s}^{-1}$. This matches the experimental expectation\cite{Iguchi1998} because for strongly correlated oxides this frequency is in the range from $10^{13}$ to $10^{14}\,\mathrm{s}^{-1}$.\cite{Phillips1989} For the low temperature process the attempt frequency is also in the order of $10^{13}\,\mathrm{s}^{-1}$.

To obtain the pre-exponential factor $A$ (Equation \ref{eq_A}), we need the averaged nearest-neighbor Mn-Mn distance in the temperature region from 310 to 400\,K, which is $d=2.93\,\mathrm{\AA}$. Moreover, $n$ equals the number of e$_\mathrm{g}$ electrons per LiMn$_2$O$_4$ unit cell, which is 8, divided by the cell volume with an averaged cubic lattice constant of 8.27\,$\mathrm{\AA}$ in the investigated temperature regime. With these data the pre-exponential factor $A$ is predicted to be $2\cdot10^5\,\Upomega^{-1}\,\mathrm{cm}^{-1}\,\mathrm{K}$.

With respect to the experiment\cite{Shimakawa1997} the conductivity is overestimated by our model by two orders of magnitude, while it should be noted that the experimentally measured conductivity changes by about four orders of magnitude between 250 and 400\,K. The overestimation is probably the consequence of the neglected correlation of the electrons which leads to a lower diffusion coefficient as predicted by Equation \ref{eq_D}. Due to occupied sites the reverse motion is more probable than another direction because the former site of the e$_\mathrm{g}$ electron is always empty shortly after the hop. This reduces the diffusion coefficient. Still, the activation energy of the hopping process can be determined precisely.

For the smaller $2\times2\times2$ supercell the phase transition is not as pronounced as for the $3\times3\times3$ supercell. If the hopping frequency data of the $2\times2\times2$ and the $3\times3\times3$ supercells are compared, it can be observed that they match above about 340\,K ($T^{-1}\approx2.94\cdot10^{-3}\,\mathrm{K}^{-1}$) which is in agreement with the results of the order parameter (Figure \ref{fig_order_parameter}). 

In conclusion, large simulation cells and nanosecond simulation times are required for an accurate identification of the phase transition based on e$_\mathrm{g}$ electron hopping. The efficiency of the HDNN method for the prediction of oxidation and spin states fulfills these requirements enabling investigations of the electronic dynamics.

\section{Conclusion}

In this work we have trained a HDNN to the atomic spins of the redox-active Mn ions in Li$_x$Mn$_2$O$_4$, which could be reproduced with an error of only about 0.03\,$\hbar$ with respect to the underlying DFT reference data. Using this HDNN we have been able to predict the oxidation and spin states of the Mn ions in MD simulations obtained using a HDNNP potential energy surface. The results confirmed that the MD simulations in our separate study\cite{Eckhoff2020a} correspond to reasonable electronic structures in which the Mn e$_\mathrm{g}$ electrons are correctly conserved and the number of Jahn-Teller distorted Mn$^\mathrm{III}$O$_6$ octahedra has been predicted reliably for different Li loadings. 

Further, the HDNN enables investigations of a charge ordering transition and electron hopping, which are not possible with conventional machine learning potentials. The charge order parameter reveals a phase transition between 280 and 300\,K. This agrees well with experimental resistivity measurements observing a transition at about 290\,K.\cite{Shimakawa1997} The hopping frequency reveals this transition between 280 and 290\,K as well. The predicted activation barrier of the thermally activated conduction process for the high temperature phase of 0.18\,eV deviates only 0.02\,eV from experimental data.\cite{Shimakawa1997} 

In conclusion, we have demonstrated that machine learning is able to provide an accurate representation of both, the structural and the electronic dynamics of Li$_x$Mn$_2$O$_4$. The investigation of the phase transition requires time and length scales that are not accessible by first-principles calculations. Therefore, the HDNN is a promising new method for the theoretical description, for example, of electrical transport properties of battery materials. Moreover, the HDNN method can be applied for molecular transition metal complexes as well. In these systems high-spin and low-spin states can occur, which lead to very different metal-ligand distances. Since for a given distance either the high-spin or the low-spin state is the electronic ground state, our method can easily identify the spin states of these systems as well.

\section*{Supplementary Material}

See Supplementary Material for (I) symmetry functions, (II) neural network training details, and (III) a three-dimensional LiMn$_2$O$_4$ representation and ab initio molecular dynamics simulations.

\begin{acknowledgments}
This project was funded by the Deutsche Forschungsgemeinschaft (DFG, German Research Foundation) - 217133147/SFB 1073, project C03. We gratefully acknowledge the funding of this project by computing time provided by the Paderborn Center for Parallel Computing (PC$^2$) and by the DFG project INST186/1294-1 FUGG (Project No.\ 405832858). J.B.\ is grateful for a DFG Heisenberg professorship BE3264/11-2 (Project No.\ 329898176).
\end{acknowledgments}

\section*{Data Availability}

The data that support the findings of this study are available from the corresponding author upon reasonable request.

\section*{References}

\bibliography{bibliography}

%aipnum4-2.bst 2019-01-14 (MD) hand-edited version of apsrev4-1.bst
%Control: key (0)
%Control: author (8) initials jnrlst
%Control: editor formatted (1) identically to author
%Control: production of article title (0) allowed
%Control: page (1) range
%Control: year (1) truncated
%Control: production of eprint (0) enabled
\begin{thebibliography}{87}%
\makeatletter
\providecommand \@ifxundefined [1]{%
 \@ifx{#1\undefined}
}%
\providecommand \@ifnum [1]{%
 \ifnum #1\expandafter \@firstoftwo
 \else \expandafter \@secondoftwo
 \fi
}%
\providecommand \@ifx [1]{%
 \ifx #1\expandafter \@firstoftwo
 \else \expandafter \@secondoftwo
 \fi
}%
\providecommand \natexlab [1]{#1}%
\providecommand \enquote  [1]{``#1''}%
\providecommand \bibnamefont  [1]{#1}%
\providecommand \bibfnamefont [1]{#1}%
\providecommand \citenamefont [1]{#1}%
\providecommand \href@noop [0]{\@secondoftwo}%
\providecommand \href [0]{\begingroup \@sanitize@url \@href}%
\providecommand \@href[1]{\@@startlink{#1}\@@href}%
\providecommand \@@href[1]{\endgroup#1\@@endlink}%
\providecommand \@sanitize@url [0]{\catcode `\\12\catcode `\$12\catcode
  `\&12\catcode `\#12\catcode `\^12\catcode `\_12\catcode `\%12\relax}%
\providecommand \@@startlink[1]{}%
\providecommand \@@endlink[0]{}%
\providecommand \url  [0]{\begingroup\@sanitize@url \@url }%
\providecommand \@url [1]{\endgroup\@href {#1}{\urlprefix }}%
\providecommand \urlprefix  [0]{URL }%
\providecommand \Eprint [0]{\href }%
\providecommand \doibase [0]{https://doi.org/}%
\providecommand \selectlanguage [0]{\@gobble}%
\providecommand \bibinfo  [0]{\@secondoftwo}%
\providecommand \bibfield  [0]{\@secondoftwo}%
\providecommand \translation [1]{[#1]}%
\providecommand \BibitemOpen [0]{}%
\providecommand \bibitemStop [0]{}%
\providecommand \bibitemNoStop [0]{.\EOS\space}%
\providecommand \EOS [0]{\spacefactor3000\relax}%
\providecommand \BibitemShut  [1]{\csname bibitem#1\endcsname}%
\let\auto@bib@innerbib\@empty
%</preamble>
\bibitem [{\citenamefont {Tarascon}\ and\ \citenamefont
  {Armand}(2001)}]{Tarascon2001}%
  \BibitemOpen
  \bibfield  {author} {\bibinfo {author} {\bibfnamefont {J.-M.}\ \bibnamefont
  {Tarascon}}\ and\ \bibinfo {author} {\bibfnamefont {M.}~\bibnamefont
  {Armand}},\ }\bibfield  {title} {\enquote {\bibinfo {title} {Issues and
  challenges facing rechargeable lithium batteries},}\ }\href@noop {}
  {\bibfield  {journal} {\bibinfo  {journal} {Nature}\ }\textbf {\bibinfo
  {volume} {414}},\ \bibinfo {pages} {359--367} (\bibinfo {year}
  {2001})}\BibitemShut {NoStop}%
\bibitem [{\citenamefont {Goodenough}\ and\ \citenamefont
  {Kim}(2010)}]{Goodenough2010}%
  \BibitemOpen
  \bibfield  {author} {\bibinfo {author} {\bibfnamefont {J.~B.}\ \bibnamefont
  {Goodenough}}\ and\ \bibinfo {author} {\bibfnamefont {Y.}~\bibnamefont
  {Kim}},\ }\bibfield  {title} {\enquote {\bibinfo {title} {Challenges for
  rechargeable {Li} batteries},}\ }\href@noop {} {\bibfield  {journal}
  {\bibinfo  {journal} {Chem. Mater.}\ }\textbf {\bibinfo {volume} {22}},\
  \bibinfo {pages} {587--603} (\bibinfo {year} {2010})}\BibitemShut {NoStop}%
\bibitem [{\citenamefont {Hautier}\ \emph {et~al.}(2010)\citenamefont
  {Hautier}, \citenamefont {Fischer}, \citenamefont {Jain}, \citenamefont
  {Mueller},\ and\ \citenamefont {Ceder}}]{Hautier2010}%
  \BibitemOpen
  \bibfield  {author} {\bibinfo {author} {\bibfnamefont {G.}~\bibnamefont
  {Hautier}}, \bibinfo {author} {\bibfnamefont {C.~C.}\ \bibnamefont
  {Fischer}}, \bibinfo {author} {\bibfnamefont {A.}~\bibnamefont {Jain}},
  \bibinfo {author} {\bibfnamefont {T.}~\bibnamefont {Mueller}},\ and\ \bibinfo
  {author} {\bibfnamefont {G.}~\bibnamefont {Ceder}},\ }\bibfield  {title}
  {\enquote {\bibinfo {title} {Finding nature's missing ternary oxide compounds
  using machine learning and density functional theory},}\ }\href@noop {}
  {\bibfield  {journal} {\bibinfo  {journal} {Chem. Mater.}\ }\textbf {\bibinfo
  {volume} {22}},\ \bibinfo {pages} {3762--3767} (\bibinfo {year}
  {2010})}\BibitemShut {NoStop}%
\bibitem [{\citenamefont {Sanchez-Lengeling}\ and\ \citenamefont
  {Aspuru-Guzik}(2018)}]{Sanchez-Lengeling2018}%
  \BibitemOpen
  \bibfield  {author} {\bibinfo {author} {\bibfnamefont {B.}~\bibnamefont
  {Sanchez-Lengeling}}\ and\ \bibinfo {author} {\bibfnamefont {A.}~\bibnamefont
  {Aspuru-Guzik}},\ }\bibfield  {title} {\enquote {\bibinfo {title} {Inverse
  molecular design using machine learning: Generative models for matter
  engineering},}\ }\href@noop {} {\bibfield  {journal} {\bibinfo  {journal}
  {Science}\ }\textbf {\bibinfo {volume} {361}},\ \bibinfo {pages} {360--365}
  (\bibinfo {year} {2018})}\BibitemShut {NoStop}%
\bibitem [{\citenamefont {Nitta}\ \emph {et~al.}(2015)\citenamefont {Nitta},
  \citenamefont {Wu}, \citenamefont {Lee},\ and\ \citenamefont
  {Yushin}}]{Nitta2015}%
  \BibitemOpen
  \bibfield  {author} {\bibinfo {author} {\bibfnamefont {N.}~\bibnamefont
  {Nitta}}, \bibinfo {author} {\bibfnamefont {F.}~\bibnamefont {Wu}}, \bibinfo
  {author} {\bibfnamefont {J.~T.}\ \bibnamefont {Lee}},\ and\ \bibinfo {author}
  {\bibfnamefont {G.}~\bibnamefont {Yushin}},\ }\bibfield  {title} {\enquote
  {\bibinfo {title} {Li-ion battery materials: present and future},}\
  }\href@noop {} {\bibfield  {journal} {\bibinfo  {journal} {Mater. Today}\
  }\textbf {\bibinfo {volume} {18}},\ \bibinfo {pages} {252--264} (\bibinfo
  {year} {2015})}\BibitemShut {NoStop}%
\bibitem [{\citenamefont {Eckhoff}, \citenamefont {Bl\"ochl},\ and\
  \citenamefont {Behler}(2020)}]{Eckhoff2020}%
  \BibitemOpen
  \bibfield  {author} {\bibinfo {author} {\bibfnamefont {M.}~\bibnamefont
  {Eckhoff}}, \bibinfo {author} {\bibfnamefont {P.~E.}\ \bibnamefont
  {Bl\"ochl}},\ and\ \bibinfo {author} {\bibfnamefont {J.}~\bibnamefont
  {Behler}},\ }\bibfield  {title} {\enquote {\bibinfo {title} {Hybrid density
  functional theory benchmark study on lithium manganese oxides},}\ }\href@noop
  {} {\bibfield  {journal} {\bibinfo  {journal} {Phys. Rev. B}\ }\textbf
  {\bibinfo {volume} {101}},\ \bibinfo {pages} {205113} (\bibinfo {year}
  {2020})}\BibitemShut {NoStop}%
\bibitem [{\citenamefont {Thackeray}\ \emph {et~al.}(1983)\citenamefont
  {Thackeray}, \citenamefont {David}, \citenamefont {Bruce},\ and\
  \citenamefont {Goodenough}}]{Thackeray1983}%
  \BibitemOpen
  \bibfield  {author} {\bibinfo {author} {\bibfnamefont {M.~M.}\ \bibnamefont
  {Thackeray}}, \bibinfo {author} {\bibfnamefont {W.~I.~F.}\ \bibnamefont
  {David}}, \bibinfo {author} {\bibfnamefont {P.~G.}\ \bibnamefont {Bruce}},\
  and\ \bibinfo {author} {\bibfnamefont {J.~B.}\ \bibnamefont {Goodenough}},\
  }\bibfield  {title} {\enquote {\bibinfo {title} {Lithium insertion into
  manganese spinels},}\ }\href@noop {} {\bibfield  {journal} {\bibinfo
  {journal} {Mat. Res. Bull.}\ }\textbf {\bibinfo {volume} {18}},\ \bibinfo
  {pages} {461--472} (\bibinfo {year} {1983})}\BibitemShut {NoStop}%
\bibitem [{\citenamefont {Thackeray}(1997)}]{Thackeray1997}%
  \BibitemOpen
  \bibfield  {author} {\bibinfo {author} {\bibfnamefont {M.~M.}\ \bibnamefont
  {Thackeray}},\ }\bibfield  {title} {\enquote {\bibinfo {title} {Manganese
  oxides for lithium batteries},}\ }\href@noop {} {\bibfield  {journal}
  {\bibinfo  {journal} {Prog. Solid State Chem.}\ }\textbf {\bibinfo {volume}
  {25}},\ \bibinfo {pages} {1--71} (\bibinfo {year} {1997})}\BibitemShut
  {NoStop}%
\bibitem [{\citenamefont {Jahn}\ and\ \citenamefont {Teller}(1937)}]{Jahn1937}%
  \BibitemOpen
  \bibfield  {author} {\bibinfo {author} {\bibfnamefont {H.~A.}\ \bibnamefont
  {Jahn}}\ and\ \bibinfo {author} {\bibfnamefont {E.}~\bibnamefont {Teller}},\
  }\bibfield  {title} {\enquote {\bibinfo {title} {Stability of polyatomic
  molecules in degenerate electronic states},}\ }\href@noop {} {\bibfield
  {journal} {\bibinfo  {journal} {Proc. Royal Soc. Lond. A}\ }\textbf {\bibinfo
  {volume} {161}},\ \bibinfo {pages} {220--235} (\bibinfo {year}
  {1937})}\BibitemShut {NoStop}%
\bibitem [{\citenamefont {Rodr{\'i}guez-Carvajal}\ \emph
  {et~al.}(1998)\citenamefont {Rodr{\'i}guez-Carvajal}, \citenamefont {Rousse},
  \citenamefont {Masquelier},\ and\ \citenamefont
  {Hervieu}}]{Rodriguez-Carvajal1998}%
  \BibitemOpen
  \bibfield  {author} {\bibinfo {author} {\bibfnamefont {J.}~\bibnamefont
  {Rodr{\'i}guez-Carvajal}}, \bibinfo {author} {\bibfnamefont {G.}~\bibnamefont
  {Rousse}}, \bibinfo {author} {\bibfnamefont {C.}~\bibnamefont {Masquelier}},\
  and\ \bibinfo {author} {\bibfnamefont {M.}~\bibnamefont {Hervieu}},\
  }\bibfield  {title} {\enquote {\bibinfo {title} {Electronic crystallization
  in a lithium battery material: Columnar ordering of electrons and holes in
  the spinel {LiMn$_2$O$_4$}},}\ }\href@noop {} {\bibfield  {journal} {\bibinfo
   {journal} {Phys. Rev. Lett.}\ }\textbf {\bibinfo {volume} {81}},\ \bibinfo
  {pages} {4660--4663} (\bibinfo {year} {1998})}\BibitemShut {NoStop}%
\bibitem [{\citenamefont {Massarotti}\ \emph {et~al.}(1999)\citenamefont
  {Massarotti}, \citenamefont {Capsoni}, \citenamefont {Bini}, \citenamefont
  {Scardi}, \citenamefont {Leoni}, \citenamefont {Baron},\ and\ \citenamefont
  {Berg}}]{Massarotti1999}%
  \BibitemOpen
  \bibfield  {author} {\bibinfo {author} {\bibfnamefont {V.}~\bibnamefont
  {Massarotti}}, \bibinfo {author} {\bibfnamefont {D.}~\bibnamefont {Capsoni}},
  \bibinfo {author} {\bibfnamefont {M.}~\bibnamefont {Bini}}, \bibinfo {author}
  {\bibfnamefont {P.}~\bibnamefont {Scardi}}, \bibinfo {author} {\bibfnamefont
  {M.}~\bibnamefont {Leoni}}, \bibinfo {author} {\bibfnamefont
  {V.}~\bibnamefont {Baron}},\ and\ \bibinfo {author} {\bibfnamefont
  {H.}~\bibnamefont {Berg}},\ }\bibfield  {title} {\enquote {\bibinfo {title}
  {{LiMn$_2$O$_4$} low-temperature phase: synchrotron and neutron diffraction
  study},}\ }\href@noop {} {\bibfield  {journal} {\bibinfo  {journal} {J. Appl.
  Cryst.}\ }\textbf {\bibinfo {volume} {32}},\ \bibinfo {pages} {1186--1189}
  (\bibinfo {year} {1999})}\BibitemShut {NoStop}%
\bibitem [{\citenamefont {Piszora}(2004)}]{Piszora2004}%
  \BibitemOpen
  \bibfield  {author} {\bibinfo {author} {\bibfnamefont {P.}~\bibnamefont
  {Piszora}},\ }\bibfield  {title} {\enquote {\bibinfo {title} {Temperature
  dependence of the order and distribution of {Mn$^{3+}$} and {Mn$^{4+}$}
  cations in orthorhombic {LiMn$_2$O$_4$}},}\ }\href@noop {} {\bibfield
  {journal} {\bibinfo  {journal} {J. Alloy. Comp.}\ }\textbf {\bibinfo {volume}
  {382}},\ \bibinfo {pages} {112--118} (\bibinfo {year} {2004})}\BibitemShut
  {NoStop}%
\bibitem [{\citenamefont {Akimoto}\ \emph {et~al.}(2004)\citenamefont
  {Akimoto}, \citenamefont {Takahashi}, \citenamefont {Kijima},\ and\
  \citenamefont {Gotoh}}]{Akimoto2004}%
  \BibitemOpen
  \bibfield  {author} {\bibinfo {author} {\bibfnamefont {J.}~\bibnamefont
  {Akimoto}}, \bibinfo {author} {\bibfnamefont {Y.}~\bibnamefont {Takahashi}},
  \bibinfo {author} {\bibfnamefont {N.}~\bibnamefont {Kijima}},\ and\ \bibinfo
  {author} {\bibfnamefont {Y.}~\bibnamefont {Gotoh}},\ }\bibfield  {title}
  {\enquote {\bibinfo {title} {Single-crystal {X}-ray structure analysis of the
  low temperature form of {LiMn$_2$O$_4$}},}\ }\href@noop {} {\bibfield
  {journal} {\bibinfo  {journal} {Solid State Ion.}\ }\textbf {\bibinfo
  {volume} {172}},\ \bibinfo {pages} {491--494} (\bibinfo {year}
  {2004})}\BibitemShut {NoStop}%
\bibitem [{\citenamefont {Sch\"utte}, \citenamefont {Colsmann},\ and\
  \citenamefont {Reuter}(1979)}]{Schuette1979}%
  \BibitemOpen
  \bibfield  {author} {\bibinfo {author} {\bibfnamefont {L.}~\bibnamefont
  {Sch\"utte}}, \bibinfo {author} {\bibfnamefont {G.}~\bibnamefont
  {Colsmann}},\ and\ \bibinfo {author} {\bibfnamefont {B.}~\bibnamefont
  {Reuter}},\ }\bibfield  {title} {\enquote {\bibinfo {title}
  {Kristallographische, elektronische und magnetische {Eigenschaften} des
  {Spinells} {Li[Mn$_2$]O$_4$}},}\ }\href@noop {} {\bibfield  {journal}
  {\bibinfo  {journal} {J. Solid State Chem.}\ }\textbf {\bibinfo {volume}
  {27}},\ \bibinfo {pages} {227--231} (\bibinfo {year} {1979})}\BibitemShut
  {NoStop}%
\bibitem [{\citenamefont {Goodenough}, \citenamefont {Manthiram},\ and\
  \citenamefont {Wnetrzewski}(1993)}]{Goodenough1993}%
  \BibitemOpen
  \bibfield  {author} {\bibinfo {author} {\bibfnamefont {J.~B.}\ \bibnamefont
  {Goodenough}}, \bibinfo {author} {\bibfnamefont {A.}~\bibnamefont
  {Manthiram}},\ and\ \bibinfo {author} {\bibfnamefont {B.}~\bibnamefont
  {Wnetrzewski}},\ }\bibfield  {title} {\enquote {\bibinfo {title} {Electrodes
  for lithium batteries},}\ }\href@noop {} {\bibfield  {journal} {\bibinfo
  {journal} {J. Power Sources}\ }\textbf {\bibinfo {volume} {43}},\ \bibinfo
  {pages} {269--275} (\bibinfo {year} {1993})}\BibitemShut {NoStop}%
\bibitem [{\citenamefont {Shimakawa}, \citenamefont {Numata},\ and\
  \citenamefont {Tabuchi}(1997)}]{Shimakawa1997}%
  \BibitemOpen
  \bibfield  {author} {\bibinfo {author} {\bibfnamefont {Y.}~\bibnamefont
  {Shimakawa}}, \bibinfo {author} {\bibfnamefont {T.}~\bibnamefont {Numata}},\
  and\ \bibinfo {author} {\bibfnamefont {J.}~\bibnamefont {Tabuchi}},\
  }\bibfield  {title} {\enquote {\bibinfo {title} {Verwey-type transition and
  magnetic properties of the {LiMn$_2$O$_4$} spinels},}\ }\href@noop {}
  {\bibfield  {journal} {\bibinfo  {journal} {J. Solid State Chem.}\ }\textbf
  {\bibinfo {volume} {131}},\ \bibinfo {pages} {138--143} (\bibinfo {year}
  {1997})}\BibitemShut {NoStop}%
\bibitem [{\citenamefont {Iguchi}, \citenamefont {Nakamura},\ and\
  \citenamefont {Aoki}(1998)}]{Iguchi1998}%
  \BibitemOpen
  \bibfield  {author} {\bibinfo {author} {\bibfnamefont {E.}~\bibnamefont
  {Iguchi}}, \bibinfo {author} {\bibfnamefont {N.}~\bibnamefont {Nakamura}},\
  and\ \bibinfo {author} {\bibfnamefont {A.}~\bibnamefont {Aoki}},\ }\bibfield
  {title} {\enquote {\bibinfo {title} {Electrical transport properties in
  {LiMn$_2$O$_4$}},}\ }\href@noop {} {\bibfield  {journal} {\bibinfo  {journal}
  {Philos. Mag. B}\ }\textbf {\bibinfo {volume} {78}},\ \bibinfo {pages}
  {65--77} (\bibinfo {year} {1998})}\BibitemShut {NoStop}%
\bibitem [{\citenamefont {Eckhoff}\ \emph {et~al.}(2020)\citenamefont
  {Eckhoff}, \citenamefont {Sch\"onewald}, \citenamefont {Risch}, \citenamefont
  {Volkert}, \citenamefont {Bl\"ochl},\ and\ \citenamefont
  {Behler}}]{Eckhoff2020a}%
  \BibitemOpen
  \bibfield  {author} {\bibinfo {author} {\bibfnamefont {M.}~\bibnamefont
  {Eckhoff}}, \bibinfo {author} {\bibfnamefont {F.}~\bibnamefont
  {Sch\"onewald}}, \bibinfo {author} {\bibfnamefont {M.}~\bibnamefont {Risch}},
  \bibinfo {author} {\bibfnamefont {C.~A.}\ \bibnamefont {Volkert}}, \bibinfo
  {author} {\bibfnamefont {P.~E.}\ \bibnamefont {Bl\"ochl}},\ and\ \bibinfo
  {author} {\bibfnamefont {J.}~\bibnamefont {Behler}},\ }\href@noop {}
  {\enquote {\bibinfo {title} {Closing the gap between theory and experiment
  for lithium manganese oxide spinels using a high-dimensional neural network
  potential},}\ } (\bibinfo {year} {2020}),\ \Eprint
  {https://arxiv.org/abs/2007.00327} {arXiv:2007.00327 [cond-mat.mtrl-sci]}
  \BibitemShut {NoStop}%
\bibitem [{\citenamefont {Takahashi}\ \emph {et~al.}(2003)\citenamefont
  {Takahashi}, \citenamefont {Akimoto}, \citenamefont {Gotoh}, \citenamefont
  {Dokko}, \citenamefont {Nishizawa},\ and\ \citenamefont
  {Uchida}}]{Takahashi2003}%
  \BibitemOpen
  \bibfield  {author} {\bibinfo {author} {\bibfnamefont {Y.}~\bibnamefont
  {Takahashi}}, \bibinfo {author} {\bibfnamefont {J.}~\bibnamefont {Akimoto}},
  \bibinfo {author} {\bibfnamefont {Y.}~\bibnamefont {Gotoh}}, \bibinfo
  {author} {\bibfnamefont {K.}~\bibnamefont {Dokko}}, \bibinfo {author}
  {\bibfnamefont {M.}~\bibnamefont {Nishizawa}},\ and\ \bibinfo {author}
  {\bibfnamefont {I.}~\bibnamefont {Uchida}},\ }\bibfield  {title} {\enquote
  {\bibinfo {title} {Structure and electron density analysis of lithium
  manganese oxides by single-crystal {X}-ray diffraction},}\ }\href@noop {}
  {\bibfield  {journal} {\bibinfo  {journal} {J. Phys. Soc. Jpn.}\ }\textbf
  {\bibinfo {volume} {72}},\ \bibinfo {pages} {1483--1490} (\bibinfo {year}
  {2003})}\BibitemShut {NoStop}%
\bibitem [{\citenamefont {Akimoto}\ \emph {et~al.}(2000)\citenamefont
  {Akimoto}, \citenamefont {Takahashi}, \citenamefont {Gotoh},\ and\
  \citenamefont {Mizuta}}]{Akimoto2000}%
  \BibitemOpen
  \bibfield  {author} {\bibinfo {author} {\bibfnamefont {J.}~\bibnamefont
  {Akimoto}}, \bibinfo {author} {\bibfnamefont {Y.}~\bibnamefont {Takahashi}},
  \bibinfo {author} {\bibfnamefont {Y.}~\bibnamefont {Gotoh}},\ and\ \bibinfo
  {author} {\bibfnamefont {S.}~\bibnamefont {Mizuta}},\ }\bibfield  {title}
  {\enquote {\bibinfo {title} {Single crystal {X}-ray diffraction study of the
  spinel-type {LiMn$_2$O$_4$}},}\ }\href@noop {} {\bibfield  {journal}
  {\bibinfo  {journal} {Chem. Mater.}\ }\textbf {\bibinfo {volume} {12}},\
  \bibinfo {pages} {3246--3248} (\bibinfo {year} {2000})}\BibitemShut {NoStop}%
\bibitem [{\citenamefont {Ouyang}, \citenamefont {Shi},\ and\ \citenamefont
  {Lei}(2009)}]{Ouyang2009}%
  \BibitemOpen
  \bibfield  {author} {\bibinfo {author} {\bibfnamefont {C.~Y.}\ \bibnamefont
  {Ouyang}}, \bibinfo {author} {\bibfnamefont {S.~Q.}\ \bibnamefont {Shi}},\
  and\ \bibinfo {author} {\bibfnamefont {M.~S.}\ \bibnamefont {Lei}},\
  }\bibfield  {title} {\enquote {\bibinfo {title} {{Jahn-Teller} distortion and
  electronic structure of {LiMn$_2$O$_4$}},}\ }\href@noop {} {\bibfield
  {journal} {\bibinfo  {journal} {J. Alloy. Comp.}\ }\textbf {\bibinfo {volume}
  {474}},\ \bibinfo {pages} {370--374} (\bibinfo {year} {2009})}\BibitemShut
  {NoStop}%
\bibitem [{\citenamefont {Mosbah}, \citenamefont {Verbaere},\ and\
  \citenamefont {Tournoux}(1983)}]{Mosbah1983}%
  \BibitemOpen
  \bibfield  {author} {\bibinfo {author} {\bibfnamefont {A.}~\bibnamefont
  {Mosbah}}, \bibinfo {author} {\bibfnamefont {A.}~\bibnamefont {Verbaere}},\
  and\ \bibinfo {author} {\bibfnamefont {M.}~\bibnamefont {Tournoux}},\
  }\bibfield  {title} {\enquote {\bibinfo {title} {Phases {Li$_x$MnO$_2$}
  rattachees au type spinelle},}\ }\href@noop {} {\bibfield  {journal}
  {\bibinfo  {journal} {Mat. Res. Bull.}\ }\textbf {\bibinfo {volume} {18}},\
  \bibinfo {pages} {1375--1381} (\bibinfo {year} {1983})}\BibitemShut {NoStop}%
\bibitem [{\citenamefont {Ohzuku}, \citenamefont {Kitagawa},\ and\
  \citenamefont {Hirai}(1989)}]{Ohzuku1989}%
  \BibitemOpen
  \bibfield  {author} {\bibinfo {author} {\bibfnamefont {T.}~\bibnamefont
  {Ohzuku}}, \bibinfo {author} {\bibfnamefont {M.}~\bibnamefont {Kitagawa}},\
  and\ \bibinfo {author} {\bibfnamefont {T.}~\bibnamefont {Hirai}},\ }\bibfield
   {title} {\enquote {\bibinfo {title} {Electrochemistry of manganese dioxide
  in lithium nonaqueous cell},}\ }\href@noop {} {\bibfield  {journal} {\bibinfo
   {journal} {J. Electrochem. Soc.}\ }\textbf {\bibinfo {volume} {136}},\
  \bibinfo {pages} {3169--3174} (\bibinfo {year} {1989})}\BibitemShut {NoStop}%
\bibitem [{\citenamefont {{de Groot}}(2001)}]{deGroot2001}%
  \BibitemOpen
  \bibfield  {author} {\bibinfo {author} {\bibfnamefont {F.}~\bibnamefont {{de
  Groot}}},\ }\bibfield  {title} {\enquote {\bibinfo {title} {High-resolution
  {X}-ray emission and {X}-ray absorption spectroscopy},}\ }\href@noop {}
  {\bibfield  {journal} {\bibinfo  {journal} {Chem. Rev.}\ }\textbf {\bibinfo
  {volume} {101}},\ \bibinfo {pages} {1779--1808} (\bibinfo {year}
  {2001})}\BibitemShut {NoStop}%
\bibitem [{\citenamefont {{de Groot}}(2005)}]{deGroot2005}%
  \BibitemOpen
  \bibfield  {author} {\bibinfo {author} {\bibfnamefont {F.}~\bibnamefont {{de
  Groot}}},\ }\bibfield  {title} {\enquote {\bibinfo {title} {Multiplet effects
  in {X}-ray spectroscopy},}\ }\href@noop {} {\bibfield  {journal} {\bibinfo
  {journal} {Coord. Chem. Rev.}\ }\textbf {\bibinfo {volume} {249}},\ \bibinfo
  {pages} {31--63} (\bibinfo {year} {2005})}\BibitemShut {NoStop}%
\bibitem [{\citenamefont {{de Groot}}\ and\ \citenamefont
  {Kotani}(2008)}]{deGroot2008}%
  \BibitemOpen
  \bibfield  {author} {\bibinfo {author} {\bibfnamefont {F.}~\bibnamefont {{de
  Groot}}}\ and\ \bibinfo {author} {\bibfnamefont {A.}~\bibnamefont {Kotani}},\
  }\href@noop {} {\emph {\bibinfo {title} {Core Level Spectroscopy of
  Solids}}}\ (\bibinfo  {publisher} {CRC Press},\ \bibinfo {address} {Boca
  Raton},\ \bibinfo {year} {2008})\BibitemShut {NoStop}%
\bibitem [{\citenamefont {Varela}\ \emph {et~al.}(2009)\citenamefont {Varela},
  \citenamefont {Oxley}, \citenamefont {Luo}, \citenamefont {Tao},
  \citenamefont {Watanabe}, \citenamefont {Lupini}, \citenamefont
  {Pantelides},\ and\ \citenamefont {Pennycook}}]{Varela2009}%
  \BibitemOpen
  \bibfield  {author} {\bibinfo {author} {\bibfnamefont {M.}~\bibnamefont
  {Varela}}, \bibinfo {author} {\bibfnamefont {M.~P.}\ \bibnamefont {Oxley}},
  \bibinfo {author} {\bibfnamefont {W.}~\bibnamefont {Luo}}, \bibinfo {author}
  {\bibfnamefont {J.}~\bibnamefont {Tao}}, \bibinfo {author} {\bibfnamefont
  {M.}~\bibnamefont {Watanabe}}, \bibinfo {author} {\bibfnamefont {A.~R.}\
  \bibnamefont {Lupini}}, \bibinfo {author} {\bibfnamefont {S.~T.}\
  \bibnamefont {Pantelides}},\ and\ \bibinfo {author} {\bibfnamefont {S.~J.}\
  \bibnamefont {Pennycook}},\ }\bibfield  {title} {\enquote {\bibinfo {title}
  {Atomic-resolution imaging of oxidation states in manganites},}\ }\href@noop
  {} {\bibfield  {journal} {\bibinfo  {journal} {Phys. Rev. B}\ }\textbf
  {\bibinfo {volume} {79}},\ \bibinfo {pages} {085117} (\bibinfo {year}
  {2009})}\BibitemShut {NoStop}%
\bibitem [{\citenamefont {Zhang}\ \emph {et~al.}(2010)\citenamefont {Zhang},
  \citenamefont {Livi}, \citenamefont {Gaillot}, \citenamefont {Stone},\ and\
  \citenamefont {Veblen}}]{Zhang2010}%
  \BibitemOpen
  \bibfield  {author} {\bibinfo {author} {\bibfnamefont {S.}~\bibnamefont
  {Zhang}}, \bibinfo {author} {\bibfnamefont {K.~J.~T.}\ \bibnamefont {Livi}},
  \bibinfo {author} {\bibfnamefont {A.-C.}\ \bibnamefont {Gaillot}}, \bibinfo
  {author} {\bibfnamefont {A.~T.}\ \bibnamefont {Stone}},\ and\ \bibinfo
  {author} {\bibfnamefont {D.~R.}\ \bibnamefont {Veblen}},\ }\bibfield  {title}
  {\enquote {\bibinfo {title} {Determination of manganese valence states in
  {(Mn$^{3+}$, Mn$^{4+}$)} minerals by electron energy-loss spectroscopy},}\
  }\href@noop {} {\bibfield  {journal} {\bibinfo  {journal} {Am. Mineral.}\
  }\textbf {\bibinfo {volume} {95}},\ \bibinfo {pages} {1741–1746} (\bibinfo
  {year} {2010})}\BibitemShut {NoStop}%
\bibitem [{\citenamefont {Sch\"onewald}\ \emph {et~al.}(2020)\citenamefont
  {Sch\"onewald}, \citenamefont {Eckhoff}, \citenamefont {Baumung},
  \citenamefont {Risch}, \citenamefont {Bl\"ochl}, \citenamefont {Behler},\
  and\ \citenamefont {Volkert}}]{Schoenewald2020}%
  \BibitemOpen
  \bibfield  {author} {\bibinfo {author} {\bibfnamefont {F.}~\bibnamefont
  {Sch\"onewald}}, \bibinfo {author} {\bibfnamefont {M.}~\bibnamefont
  {Eckhoff}}, \bibinfo {author} {\bibfnamefont {M.}~\bibnamefont {Baumung}},
  \bibinfo {author} {\bibfnamefont {M.}~\bibnamefont {Risch}}, \bibinfo
  {author} {\bibfnamefont {P.~E.}\ \bibnamefont {Bl\"ochl}}, \bibinfo {author}
  {\bibfnamefont {J.}~\bibnamefont {Behler}},\ and\ \bibinfo {author}
  {\bibfnamefont {C.~A.}\ \bibnamefont {Volkert}},\ }\href@noop {} {\enquote
  {\bibinfo {title} {A criticial view on e$_\mathrm{g}$ occupancy as a
  descriptor for oxygen evolution catalytic activity in {LiMn$_2$O$_4$}
  nanoparticles},}\ } (\bibinfo {year} {2020}),\ \Eprint
  {https://arxiv.org/abs/2007.04217} {arXiv:2007.04217 [cond-mat.mtrl-sci]}
  \BibitemShut {NoStop}%
\bibitem [{\citenamefont {Behler}(2016)}]{Behler2016}%
  \BibitemOpen
  \bibfield  {author} {\bibinfo {author} {\bibfnamefont {J.}~\bibnamefont
  {Behler}},\ }\bibfield  {title} {\enquote {\bibinfo {title} {Perspective:
  Machine learning potentials for atomistic simulations},}\ }\href@noop {}
  {\bibfield  {journal} {\bibinfo  {journal} {J. Chem. Phys.}\ }\textbf
  {\bibinfo {volume} {145}},\ \bibinfo {pages} {170901} (\bibinfo {year}
  {2016})}\BibitemShut {NoStop}%
\bibitem [{\citenamefont {Deringer}, \citenamefont {Caro},\ and\ \citenamefont
  {Csanyi}(2019)}]{P5673}%
  \BibitemOpen
  \bibfield  {author} {\bibinfo {author} {\bibfnamefont {V.~L.}\ \bibnamefont
  {Deringer}}, \bibinfo {author} {\bibfnamefont {M.~A.}\ \bibnamefont {Caro}},\
  and\ \bibinfo {author} {\bibfnamefont {G.}~\bibnamefont {Csanyi}},\
  }\bibfield  {title} {\enquote {\bibinfo {title} {Machine learning interatomic
  potentials as emerging tools for materials science},}\ }\href@noop {}
  {\bibfield  {journal} {\bibinfo  {journal} {Adv. Mater.}\ }\textbf {\bibinfo
  {volume} {31}},\ \bibinfo {pages} {1902765} (\bibinfo {year}
  {2019})}\BibitemShut {NoStop}%
\bibitem [{\citenamefont {Noé}\ \emph {et~al.}(2020)\citenamefont {Noé},
  \citenamefont {Tkatchenko}, \citenamefont {M\"uller},\ and\ \citenamefont
  {Clementi}}]{P5793}%
  \BibitemOpen
  \bibfield  {author} {\bibinfo {author} {\bibfnamefont {F.}~\bibnamefont
  {Noé}}, \bibinfo {author} {\bibfnamefont {A.}~\bibnamefont {Tkatchenko}},
  \bibinfo {author} {\bibfnamefont {K.-R.}\ \bibnamefont {M\"uller}},\ and\
  \bibinfo {author} {\bibfnamefont {C.}~\bibnamefont {Clementi}},\ }\bibfield
  {title} {\enquote {\bibinfo {title} {Machine learning for molecular
  simulation},}\ }\href@noop {} {\bibfield  {journal} {\bibinfo  {journal}
  {Ann. Rev. Phys. Chem.}\ }\textbf {\bibinfo {volume} {71}},\ \bibinfo {pages}
  {361--390} (\bibinfo {year} {2020})}\BibitemShut {NoStop}%
\bibitem [{\citenamefont {Behler}\ and\ \citenamefont
  {Parrinello}(2007)}]{Behler2007}%
  \BibitemOpen
  \bibfield  {author} {\bibinfo {author} {\bibfnamefont {J.}~\bibnamefont
  {Behler}}\ and\ \bibinfo {author} {\bibfnamefont {M.}~\bibnamefont
  {Parrinello}},\ }\bibfield  {title} {\enquote {\bibinfo {title} {Generalized
  neural-network representation of high-dimensional potential-energy
  surfaces},}\ }\href@noop {} {\bibfield  {journal} {\bibinfo  {journal} {Phys.
  Rev. Lett.}\ }\textbf {\bibinfo {volume} {98}},\ \bibinfo {pages} {146401}
  (\bibinfo {year} {2007})}\BibitemShut {NoStop}%
\bibitem [{\citenamefont {Behler}(2011)}]{Behler2011}%
  \BibitemOpen
  \bibfield  {author} {\bibinfo {author} {\bibfnamefont {J.}~\bibnamefont
  {Behler}},\ }\bibfield  {title} {\enquote {\bibinfo {title} {Atom-centered
  symmetry functions for constructing high-dimensional neural network
  potentials},}\ }\href@noop {} {\bibfield  {journal} {\bibinfo  {journal} {J.
  Chem. Phys.}\ }\textbf {\bibinfo {volume} {134}},\ \bibinfo {pages} {074106}
  (\bibinfo {year} {2011})}\BibitemShut {NoStop}%
\bibitem [{\citenamefont {Behler}(2017)}]{Behler2017}%
  \BibitemOpen
  \bibfield  {author} {\bibinfo {author} {\bibfnamefont {J.}~\bibnamefont
  {Behler}},\ }\bibfield  {title} {\enquote {\bibinfo {title} {First principles
  neural network potentials for reactive simulations of large molecular and
  condensed systems},}\ }\href@noop {} {\bibfield  {journal} {\bibinfo
  {journal} {Angew. Chem. Int. Ed.}\ }\textbf {\bibinfo {volume} {56}},\
  \bibinfo {pages} {12828--12840} (\bibinfo {year} {2017})}\BibitemShut
  {NoStop}%
\bibitem [{\citenamefont {Behler}\ \emph {et~al.}(2008)\citenamefont {Behler},
  \citenamefont {Marto\v{n}\'{a}k}, \citenamefont {Donadio},\ and\
  \citenamefont {Parrinello}}]{Behler2008}%
  \BibitemOpen
  \bibfield  {author} {\bibinfo {author} {\bibfnamefont {J.}~\bibnamefont
  {Behler}}, \bibinfo {author} {\bibfnamefont {R.}~\bibnamefont
  {Marto\v{n}\'{a}k}}, \bibinfo {author} {\bibfnamefont {D.}~\bibnamefont
  {Donadio}},\ and\ \bibinfo {author} {\bibfnamefont {M.}~\bibnamefont
  {Parrinello}},\ }\bibfield  {title} {\enquote {\bibinfo {title} {Metadynamics
  simulations of the high-pressure phases of silicon employing a
  high-dimensional neural network potential},}\ }\href@noop {} {\bibfield
  {journal} {\bibinfo  {journal} {Phys. Rev. Lett.}\ }\textbf {\bibinfo
  {volume} {100}},\ \bibinfo {pages} {185501} (\bibinfo {year}
  {2008})}\BibitemShut {NoStop}%
\bibitem [{\citenamefont {Artrith}, \citenamefont {Hiller},\ and\ \citenamefont
  {Behler}(2013)}]{Artrith2013}%
  \BibitemOpen
  \bibfield  {author} {\bibinfo {author} {\bibfnamefont {N.}~\bibnamefont
  {Artrith}}, \bibinfo {author} {\bibfnamefont {B.}~\bibnamefont {Hiller}},\
  and\ \bibinfo {author} {\bibfnamefont {J.}~\bibnamefont {Behler}},\
  }\bibfield  {title} {\enquote {\bibinfo {title} {Neural network potentials
  for metals and oxides - first applications to copper clusters at zinc
  oxide},}\ }\href@noop {} {\bibfield  {journal} {\bibinfo  {journal} {Phys.
  Status Solidi B}\ }\textbf {\bibinfo {volume} {250}},\ \bibinfo {pages}
  {1191--1203} (\bibinfo {year} {2013})}\BibitemShut {NoStop}%
\bibitem [{\citenamefont {Gastegger}\ and\ \citenamefont
  {Marquetand}(2015)}]{Gastegger2015}%
  \BibitemOpen
  \bibfield  {author} {\bibinfo {author} {\bibfnamefont {M.}~\bibnamefont
  {Gastegger}}\ and\ \bibinfo {author} {\bibfnamefont {P.}~\bibnamefont
  {Marquetand}},\ }\bibfield  {title} {\enquote {\bibinfo {title}
  {High-dimensional neural network potentials for organic reactions and an
  improved training algorithm},}\ }\href@noop {} {\bibfield  {journal}
  {\bibinfo  {journal} {J. Chem. Theory Comput.}\ }\textbf {\bibinfo {volume}
  {11}},\ \bibinfo {pages} {2187--2198} (\bibinfo {year} {2015})}\BibitemShut
  {NoStop}%
\bibitem [{\citenamefont {Morawietz}\ \emph {et~al.}(2016)\citenamefont
  {Morawietz}, \citenamefont {Singraber}, \citenamefont {Dellago},\ and\
  \citenamefont {Behler}}]{Morawietz2016}%
  \BibitemOpen
  \bibfield  {author} {\bibinfo {author} {\bibfnamefont {T.}~\bibnamefont
  {Morawietz}}, \bibinfo {author} {\bibfnamefont {A.}~\bibnamefont
  {Singraber}}, \bibinfo {author} {\bibfnamefont {C.}~\bibnamefont {Dellago}},\
  and\ \bibinfo {author} {\bibfnamefont {J.}~\bibnamefont {Behler}},\
  }\bibfield  {title} {\enquote {\bibinfo {title} {How van der {Waals}
  interactions determine the unique properties of water},}\ }\href@noop {}
  {\bibfield  {journal} {\bibinfo  {journal} {PNAS}\ }\textbf {\bibinfo
  {volume} {113}},\ \bibinfo {pages} {8368--8373} (\bibinfo {year}
  {2016})}\BibitemShut {NoStop}%
\bibitem [{\citenamefont {Hellstr\"om}\ and\ \citenamefont
  {Behler}(2016)}]{Hellstrom2016}%
  \BibitemOpen
  \bibfield  {author} {\bibinfo {author} {\bibfnamefont {M.}~\bibnamefont
  {Hellstr\"om}}\ and\ \bibinfo {author} {\bibfnamefont {J.}~\bibnamefont
  {Behler}},\ }\bibfield  {title} {\enquote {\bibinfo {title}
  {Concentration-dependent proton transfer mechanisms in aqueous {NaOH}
  solutions: From acceptor-driven to donor-driven and back},}\ }\href@noop {}
  {\bibfield  {journal} {\bibinfo  {journal} {J. Phys. Chem. Lett.}\ }\textbf
  {\bibinfo {volume} {7}},\ \bibinfo {pages} {3302--3306} (\bibinfo {year}
  {2016})}\BibitemShut {NoStop}%
\bibitem [{\citenamefont {Natarajan}\ and\ \citenamefont
  {Behler}(2016)}]{Natarajan2016}%
  \BibitemOpen
  \bibfield  {author} {\bibinfo {author} {\bibfnamefont {S.~K.}\ \bibnamefont
  {Natarajan}}\ and\ \bibinfo {author} {\bibfnamefont {J.}~\bibnamefont
  {Behler}},\ }\bibfield  {title} {\enquote {\bibinfo {title} {Neural network
  molecular dynamics simulations of solid-liquid interfaces: water at low-index
  copper surfaces},}\ }\href@noop {} {\bibfield  {journal} {\bibinfo  {journal}
  {Phys. Chem. Chem. Phys.}\ }\textbf {\bibinfo {volume} {18}},\ \bibinfo
  {pages} {28704--28725} (\bibinfo {year} {2016})}\BibitemShut {NoStop}%
\bibitem [{\citenamefont {Eckhoff}\ and\ \citenamefont
  {Behler}(2019)}]{Eckhoff2019}%
  \BibitemOpen
  \bibfield  {author} {\bibinfo {author} {\bibfnamefont {M.}~\bibnamefont
  {Eckhoff}}\ and\ \bibinfo {author} {\bibfnamefont {J.}~\bibnamefont
  {Behler}},\ }\bibfield  {title} {\enquote {\bibinfo {title} {From molecular
  fragments to the bulk: Development of a neural network potential for
  {MOF-5}},}\ }\href@noop {} {\bibfield  {journal} {\bibinfo  {journal} {J.
  Chem. Theory Comput.}\ }\textbf {\bibinfo {volume} {15}},\ \bibinfo {pages}
  {3793--3809} (\bibinfo {year} {2019})}\BibitemShut {NoStop}%
\bibitem [{\citenamefont {Artrith}, \citenamefont {Morawietz},\ and\
  \citenamefont {Behler}(2011)}]{P2962}%
  \BibitemOpen
  \bibfield  {author} {\bibinfo {author} {\bibfnamefont {N.}~\bibnamefont
  {Artrith}}, \bibinfo {author} {\bibfnamefont {T.}~\bibnamefont {Morawietz}},\
  and\ \bibinfo {author} {\bibfnamefont {J.}~\bibnamefont {Behler}},\
  }\bibfield  {title} {\enquote {\bibinfo {title} {High-dimensional
  neural-network potentials for multicomponent systems: Applications to zinc
  oxide},}\ }\href@noop {} {\bibfield  {journal} {\bibinfo  {journal} {Phys.
  Rev. B}\ }\textbf {\bibinfo {volume} {83}},\ \bibinfo {pages} {153101}
  (\bibinfo {year} {2011})}\BibitemShut {NoStop}%
\bibitem [{\citenamefont {Ghasemi}\ \emph {et~al.}(2015)\citenamefont
  {Ghasemi}, \citenamefont {Hofstetter}, \citenamefont {Saha},\ and\
  \citenamefont {Goedecker}}]{P4419}%
  \BibitemOpen
  \bibfield  {author} {\bibinfo {author} {\bibfnamefont {S.~A.}\ \bibnamefont
  {Ghasemi}}, \bibinfo {author} {\bibfnamefont {A.}~\bibnamefont {Hofstetter}},
  \bibinfo {author} {\bibfnamefont {S.}~\bibnamefont {Saha}},\ and\ \bibinfo
  {author} {\bibfnamefont {S.}~\bibnamefont {Goedecker}},\ }\bibfield  {title}
  {\enquote {\bibinfo {title} {Interatomic potentials for ionic systems with
  density functional accuracy based on charge densities obtained by a neural
  network},}\ }\href@noop {} {\bibfield  {journal} {\bibinfo  {journal} {Phys.
  Rev. B}\ }\textbf {\bibinfo {volume} {92}},\ \bibinfo {pages} {045131}
  (\bibinfo {year} {2015})}\BibitemShut {NoStop}%
\bibitem [{\citenamefont {Houlding}, \citenamefont {Liem},\ and\ \citenamefont
  {Popelier}(2007)}]{P2391}%
  \BibitemOpen
  \bibfield  {author} {\bibinfo {author} {\bibfnamefont {S.}~\bibnamefont
  {Houlding}}, \bibinfo {author} {\bibfnamefont {S.~Y.}\ \bibnamefont {Liem}},\
  and\ \bibinfo {author} {\bibfnamefont {P.~L.~A.}\ \bibnamefont {Popelier}},\
  }\bibfield  {title} {\enquote {\bibinfo {title} {A polarizable high-rank
  quantum topological electrostatic potential developed using neural networks:
  Molecular dynamics simulations on the hydrogen fluoride dimer},}\ }\href@noop
  {} {\bibfield  {journal} {\bibinfo  {journal} {Int. J. Quantum Chem.}\
  }\textbf {\bibinfo {volume} {107}},\ \bibinfo {pages} {2817--2827} (\bibinfo
  {year} {2007})}\BibitemShut {NoStop}%
\bibitem [{\citenamefont {Rupp}\ \emph {et~al.}(2012)\citenamefont {Rupp},
  \citenamefont {Tkatchenko}, \citenamefont {M\"uller},\ and\ \citenamefont
  {{von Lilienfeld}}}]{P3136}%
  \BibitemOpen
  \bibfield  {author} {\bibinfo {author} {\bibfnamefont {M.}~\bibnamefont
  {Rupp}}, \bibinfo {author} {\bibfnamefont {A.}~\bibnamefont {Tkatchenko}},
  \bibinfo {author} {\bibfnamefont {K.-R.}\ \bibnamefont {M\"uller}},\ and\
  \bibinfo {author} {\bibfnamefont {O.~A.}\ \bibnamefont {{von Lilienfeld}}},\
  }\bibfield  {title} {\enquote {\bibinfo {title} {Fast and accurate modeling
  of molecular atomization energies with machine learning},}\ }\href@noop {}
  {\bibfield  {journal} {\bibinfo  {journal} {Phys. Rev. Lett.}\ }\textbf
  {\bibinfo {volume} {108}},\ \bibinfo {pages} {058301} (\bibinfo {year}
  {2012})}\BibitemShut {NoStop}%
\bibitem [{\citenamefont {Montavon}\ \emph {et~al.}(2013)\citenamefont
  {Montavon}, \citenamefont {Rupp}, \citenamefont {Gobre}, \citenamefont
  {Vazquez-Mayagoitia}, \citenamefont {Hansen}, \citenamefont {Tkatchenko},
  \citenamefont {Mueller},\ and\ \citenamefont {{von Lilienfeld}}}]{P4514}%
  \BibitemOpen
  \bibfield  {author} {\bibinfo {author} {\bibfnamefont {G.}~\bibnamefont
  {Montavon}}, \bibinfo {author} {\bibfnamefont {M.}~\bibnamefont {Rupp}},
  \bibinfo {author} {\bibfnamefont {V.}~\bibnamefont {Gobre}}, \bibinfo
  {author} {\bibfnamefont {A.}~\bibnamefont {Vazquez-Mayagoitia}}, \bibinfo
  {author} {\bibfnamefont {K.}~\bibnamefont {Hansen}}, \bibinfo {author}
  {\bibfnamefont {A.}~\bibnamefont {Tkatchenko}}, \bibinfo {author}
  {\bibfnamefont {K.-R.}\ \bibnamefont {Mueller}},\ and\ \bibinfo {author}
  {\bibfnamefont {O.~A.}\ \bibnamefont {{von Lilienfeld}}},\ }\bibfield
  {title} {\enquote {\bibinfo {title} {Machine learning of molecular electronic
  properties in chemical compound space},}\ }\href@noop {} {\bibfield
  {journal} {\bibinfo  {journal} {New J. Phys.}\ }\textbf {\bibinfo {volume}
  {15}},\ \bibinfo {pages} {095003} (\bibinfo {year} {2013})}\BibitemShut
  {NoStop}%
\bibitem [{\citenamefont {Li}\ \emph {et~al.}(2018)\citenamefont {Li},
  \citenamefont {Omidvar}, \citenamefont {Chin}, \citenamefont {Robb},
  \citenamefont {Morris}, \citenamefont {Achenie},\ and\ \citenamefont
  {Xin}}]{Li2018}%
  \BibitemOpen
  \bibfield  {author} {\bibinfo {author} {\bibfnamefont {Z.}~\bibnamefont
  {Li}}, \bibinfo {author} {\bibfnamefont {N.}~\bibnamefont {Omidvar}},
  \bibinfo {author} {\bibfnamefont {W.~S.}\ \bibnamefont {Chin}}, \bibinfo
  {author} {\bibfnamefont {E.}~\bibnamefont {Robb}}, \bibinfo {author}
  {\bibfnamefont {A.}~\bibnamefont {Morris}}, \bibinfo {author} {\bibfnamefont
  {L.}~\bibnamefont {Achenie}},\ and\ \bibinfo {author} {\bibfnamefont
  {H.}~\bibnamefont {Xin}},\ }\bibfield  {title} {\enquote {\bibinfo {title}
  {Machine-learning energy gaps of porphyrins with molecular graph
  representations},}\ }\href@noop {} {\bibfield  {journal} {\bibinfo  {journal}
  {J. Phys. Chem. A}\ }\textbf {\bibinfo {volume} {122}},\ \bibinfo {pages}
  {4571--4578} (\bibinfo {year} {2018})}\BibitemShut {NoStop}%
\bibitem [{\citenamefont {Janet}\ \emph {et~al.}(2020)\citenamefont {Janet},
  \citenamefont {Ramesh}, \citenamefont {Duan},\ and\ \citenamefont
  {Kulik}}]{Janet2020}%
  \BibitemOpen
  \bibfield  {author} {\bibinfo {author} {\bibfnamefont {J.~P.}\ \bibnamefont
  {Janet}}, \bibinfo {author} {\bibfnamefont {S.}~\bibnamefont {Ramesh}},
  \bibinfo {author} {\bibfnamefont {C.}~\bibnamefont {Duan}},\ and\ \bibinfo
  {author} {\bibfnamefont {H.~J.}\ \bibnamefont {Kulik}},\ }\bibfield  {title}
  {\enquote {\bibinfo {title} {Accurate multiobjective design in a space of
  millions of transition metal complexes with neural-network-driven efficient
  global optimization},}\ }\href@noop {} {\bibfield  {journal} {\bibinfo
  {journal} {ACS Cent. Sci.}\ }\textbf {\bibinfo {volume} {6}},\ \bibinfo
  {pages} {513--524} (\bibinfo {year} {2020})}\BibitemShut {NoStop}%
\bibitem [{\citenamefont {Rupp}, \citenamefont {Ramakrishnan},\ and\
  \citenamefont {{von Lilienfeld}}(2015)}]{P4478}%
  \BibitemOpen
  \bibfield  {author} {\bibinfo {author} {\bibfnamefont {M.}~\bibnamefont
  {Rupp}}, \bibinfo {author} {\bibfnamefont {R.}~\bibnamefont {Ramakrishnan}},\
  and\ \bibinfo {author} {\bibfnamefont {O.~A.}\ \bibnamefont {{von
  Lilienfeld}}},\ }\bibfield  {title} {\enquote {\bibinfo {title} {Machine
  learning for quantum mechanical properties of atoms in molecules},}\
  }\href@noop {} {\bibfield  {journal} {\bibinfo  {journal} {J. Phys. Chem.
  Lett.}\ }\textbf {\bibinfo {volume} {6}},\ \bibinfo {pages} {3309} (\bibinfo
  {year} {2015})}\BibitemShut {NoStop}%
\bibitem [{\citenamefont {Sch\"utt}\ \emph {et~al.}(2019)\citenamefont
  {Sch\"utt}, \citenamefont {Gastegger}, \citenamefont {Tkatchenko},
  \citenamefont {M\"uller},\ and\ \citenamefont {Maurer}}]{P5811}%
  \BibitemOpen
  \bibfield  {author} {\bibinfo {author} {\bibfnamefont {K.}~\bibnamefont
  {Sch\"utt}}, \bibinfo {author} {\bibfnamefont {M.}~\bibnamefont {Gastegger}},
  \bibinfo {author} {\bibfnamefont {A.}~\bibnamefont {Tkatchenko}}, \bibinfo
  {author} {\bibfnamefont {K.-R.}\ \bibnamefont {M\"uller}},\ and\ \bibinfo
  {author} {\bibfnamefont {R.}~\bibnamefont {Maurer}},\ }\bibfield  {title}
  {\enquote {\bibinfo {title} {Unifying machine learning and quantum chemistry
  with a deep neural network for molecular wavefunctions},}\ }\href@noop {}
  {\bibfield  {journal} {\bibinfo  {journal} {Nature Comm.}\ }\textbf {\bibinfo
  {volume} {10}},\ \bibinfo {pages} {5024} (\bibinfo {year}
  {2019})}\BibitemShut {NoStop}%
\bibitem [{\citenamefont {Janet}\ and\ \citenamefont
  {Kulik}(2017{\natexlab{a}})}]{Janet2017}%
  \BibitemOpen
  \bibfield  {author} {\bibinfo {author} {\bibfnamefont {J.~P.}\ \bibnamefont
  {Janet}}\ and\ \bibinfo {author} {\bibfnamefont {H.~J.}\ \bibnamefont
  {Kulik}},\ }\bibfield  {title} {\enquote {\bibinfo {title} {Predicting
  electronic structure properties of transition metal complexes with neural
  networks},}\ }\href@noop {} {\bibfield  {journal} {\bibinfo  {journal} {Chem.
  Sci.}\ }\textbf {\bibinfo {volume} {8}},\ \bibinfo {pages} {5137--5152}
  (\bibinfo {year} {2017}{\natexlab{a}})}\BibitemShut {NoStop}%
\bibitem [{\citenamefont {Janet}\ and\ \citenamefont
  {Kulik}(2017{\natexlab{b}})}]{Janet2017a}%
  \BibitemOpen
  \bibfield  {author} {\bibinfo {author} {\bibfnamefont {J.~P.}\ \bibnamefont
  {Janet}}\ and\ \bibinfo {author} {\bibfnamefont {H.~J.}\ \bibnamefont
  {Kulik}},\ }\bibfield  {title} {\enquote {\bibinfo {title} {Resolving
  transition metal chemical space: Feature selection for machine learning and
  structure–property relationships},}\ }\href@noop {} {\bibfield  {journal}
  {\bibinfo  {journal} {J. Phys. Chem. A}\ }\textbf {\bibinfo {volume} {121}},\
  \bibinfo {pages} {8939--8954} (\bibinfo {year}
  {2017}{\natexlab{b}})}\BibitemShut {NoStop}%
\bibitem [{\citenamefont {Janet}, \citenamefont {Chan},\ and\ \citenamefont
  {Kulik}(2018)}]{Janet2018}%
  \BibitemOpen
  \bibfield  {author} {\bibinfo {author} {\bibfnamefont {J.~P.}\ \bibnamefont
  {Janet}}, \bibinfo {author} {\bibfnamefont {L.}~\bibnamefont {Chan}},\ and\
  \bibinfo {author} {\bibfnamefont {H.~J.}\ \bibnamefont {Kulik}},\ }\bibfield
  {title} {\enquote {\bibinfo {title} {Accelerating chemical discovery with
  machine learning: Simulated evolution of spin crossover complexes with an
  artificial neural network},}\ }\href@noop {} {\bibfield  {journal} {\bibinfo
  {journal} {J. Phys. Chem. Lett.}\ }\textbf {\bibinfo {volume} {9}},\ \bibinfo
  {pages} {1064--1071} (\bibinfo {year} {2018})}\BibitemShut {NoStop}%
\bibitem [{\citenamefont {Taylor}\ \emph {et~al.}(2020)\citenamefont {Taylor},
  \citenamefont {Yang}, \citenamefont {Lin}, \citenamefont {Nandy},
  \citenamefont {Janet}, \citenamefont {Duan},\ and\ \citenamefont
  {Kulik}}]{Taylor2020}%
  \BibitemOpen
  \bibfield  {author} {\bibinfo {author} {\bibfnamefont {M.~G.}\ \bibnamefont
  {Taylor}}, \bibinfo {author} {\bibfnamefont {T.}~\bibnamefont {Yang}},
  \bibinfo {author} {\bibfnamefont {S.}~\bibnamefont {Lin}}, \bibinfo {author}
  {\bibfnamefont {A.}~\bibnamefont {Nandy}}, \bibinfo {author} {\bibfnamefont
  {J.~P.}\ \bibnamefont {Janet}}, \bibinfo {author} {\bibfnamefont
  {C.}~\bibnamefont {Duan}},\ and\ \bibinfo {author} {\bibfnamefont {H.~J.}\
  \bibnamefont {Kulik}},\ }\bibfield  {title} {\enquote {\bibinfo {title}
  {Seeing is believing: Experimental spin states from machine learning model
  structure predictions},}\ }\href@noop {} {\bibfield  {journal} {\bibinfo
  {journal} {J. Phys. Chem. A}\ }\textbf {\bibinfo {volume} {124}},\ \bibinfo
  {pages} {3286--3299} (\bibinfo {year} {2020})}\BibitemShut {NoStop}%
\bibitem [{\citenamefont {Sotoudeh}\ \emph {et~al.}(2017)\citenamefont
  {Sotoudeh}, \citenamefont {Rajpurohit}, \citenamefont {Bl\"ochl},
  \citenamefont {Mierwaldt}, \citenamefont {Norpoth}, \citenamefont {Roddatis},
  \citenamefont {Mildner}, \citenamefont {Kressdorf}, \citenamefont {Ifland},\
  and\ \citenamefont {Jooss}}]{Sotoudeh2017}%
  \BibitemOpen
  \bibfield  {author} {\bibinfo {author} {\bibfnamefont {M.}~\bibnamefont
  {Sotoudeh}}, \bibinfo {author} {\bibfnamefont {S.}~\bibnamefont
  {Rajpurohit}}, \bibinfo {author} {\bibfnamefont {P.}~\bibnamefont
  {Bl\"ochl}}, \bibinfo {author} {\bibfnamefont {D.}~\bibnamefont {Mierwaldt}},
  \bibinfo {author} {\bibfnamefont {J.}~\bibnamefont {Norpoth}}, \bibinfo
  {author} {\bibfnamefont {V.}~\bibnamefont {Roddatis}}, \bibinfo {author}
  {\bibfnamefont {S.}~\bibnamefont {Mildner}}, \bibinfo {author} {\bibfnamefont
  {B.}~\bibnamefont {Kressdorf}}, \bibinfo {author} {\bibfnamefont
  {B.}~\bibnamefont {Ifland}},\ and\ \bibinfo {author} {\bibfnamefont
  {C.}~\bibnamefont {Jooss}},\ }\bibfield  {title} {\enquote {\bibinfo {title}
  {Electronic structure of {Pr$_{1-x}$Ca$_x$MnO$_3$}},}\ }\href@noop {}
  {\bibfield  {journal} {\bibinfo  {journal} {Phys. Rev. B}\ }\textbf {\bibinfo
  {volume} {95}},\ \bibinfo {pages} {235150} (\bibinfo {year}
  {2017})}\BibitemShut {NoStop}%
\bibitem [{\citenamefont {{McCafferty}}(2010)}]{McCafferty2010}%
  \BibitemOpen
  \bibfield  {author} {\bibinfo {author} {\bibfnamefont {E.}~\bibnamefont
  {{McCafferty}}},\ }\href@noop {} {\emph {\bibinfo {title} {Introduction to
  Corrosion Science}}}\ (\bibinfo  {publisher} {Springer},\ \bibinfo {address}
  {New York},\ \bibinfo {year} {2010})\BibitemShut {NoStop}%
\bibitem [{\citenamefont {Perutz}\ \emph {et~al.}(1998)\citenamefont {Perutz},
  \citenamefont {Wilkinson}, \citenamefont {Paoli},\ and\ \citenamefont
  {Dodson}}]{Perutz1998}%
  \BibitemOpen
  \bibfield  {author} {\bibinfo {author} {\bibfnamefont {M.~F.}\ \bibnamefont
  {Perutz}}, \bibinfo {author} {\bibfnamefont {A.~J.}\ \bibnamefont
  {Wilkinson}}, \bibinfo {author} {\bibfnamefont {M.}~\bibnamefont {Paoli}},\
  and\ \bibinfo {author} {\bibfnamefont {G.~G.}\ \bibnamefont {Dodson}},\
  }\bibfield  {title} {\enquote {\bibinfo {title} {The stereochemical mechanism
  of the cooperative effects in hemoglobin revisited},}\ }\href@noop {}
  {\bibfield  {journal} {\bibinfo  {journal} {Annu. Rev. Biophys. Biomol.
  Struct.}\ }\textbf {\bibinfo {volume} {27}},\ \bibinfo {pages} {1--34}
  (\bibinfo {year} {1998})}\BibitemShut {NoStop}%
\bibitem [{\citenamefont {Morawietz}, \citenamefont {Sharma},\ and\
  \citenamefont {Behler}(2012)}]{P3132}%
  \BibitemOpen
  \bibfield  {author} {\bibinfo {author} {\bibfnamefont {T.}~\bibnamefont
  {Morawietz}}, \bibinfo {author} {\bibfnamefont {V.}~\bibnamefont {Sharma}},\
  and\ \bibinfo {author} {\bibfnamefont {J.}~\bibnamefont {Behler}},\
  }\bibfield  {title} {\enquote {\bibinfo {title} {A neural network
  potential-energy surface for the water dimer based on environment-dependent
  atomic energies and charges},}\ }\href@noop {} {\bibfield  {journal}
  {\bibinfo  {journal} {J. Chem. Phys.}\ }\textbf {\bibinfo {volume} {136}},\
  \bibinfo {pages} {064103} (\bibinfo {year} {2012})}\BibitemShut {NoStop}%
\bibitem [{\citenamefont {Hornik}, \citenamefont {Stinchcombe},\ and\
  \citenamefont {White}(1989)}]{Hornik1989}%
  \BibitemOpen
  \bibfield  {author} {\bibinfo {author} {\bibfnamefont {K.}~\bibnamefont
  {Hornik}}, \bibinfo {author} {\bibfnamefont {M.}~\bibnamefont
  {Stinchcombe}},\ and\ \bibinfo {author} {\bibfnamefont {H.}~\bibnamefont
  {White}},\ }\bibfield  {title} {\enquote {\bibinfo {title} {Multilayer
  feedforward networks are universal approximators},}\ }\href@noop {}
  {\bibfield  {journal} {\bibinfo  {journal} {Neural Netw.}\ }\textbf {\bibinfo
  {volume} {2}},\ \bibinfo {pages} {359--366} (\bibinfo {year}
  {1989})}\BibitemShut {NoStop}%
\bibitem [{\citenamefont {Behler}(2015)}]{Behler2015}%
  \BibitemOpen
  \bibfield  {author} {\bibinfo {author} {\bibfnamefont {J.}~\bibnamefont
  {Behler}},\ }\bibfield  {title} {\enquote {\bibinfo {title} {Constructing
  high-dimensional neural network potentials: A tutorial review},}\ }\href@noop
  {} {\bibfield  {journal} {\bibinfo  {journal} {Int. J. Quantum Chem.}\
  }\textbf {\bibinfo {volume} {115}},\ \bibinfo {pages} {1032--1050} (\bibinfo
  {year} {2015})}\BibitemShut {NoStop}%
\bibitem [{\citenamefont {Bl\"ochl}(1994)}]{Bloechl1994}%
  \BibitemOpen
  \bibfield  {author} {\bibinfo {author} {\bibfnamefont {P.~E.}\ \bibnamefont
  {Bl\"ochl}},\ }\bibfield  {title} {\enquote {\bibinfo {title} {Projector
  augmented-wave method},}\ }\href@noop {} {\bibfield  {journal} {\bibinfo
  {journal} {Phys. Rev. B}\ }\textbf {\bibinfo {volume} {50}},\ \bibinfo
  {pages} {17953--17979} (\bibinfo {year} {1994})}\BibitemShut {NoStop}%
\bibitem [{\citenamefont {Bl\"ochl}(2016)}]{CP-PAW}%
  \BibitemOpen
  \bibfield  {author} {\bibinfo {author} {\bibfnamefont {P.~E.}\ \bibnamefont
  {Bl\"ochl}},\ }\bibfield  {title} {\enquote {\bibinfo {title} {{CP-PAW}},}\
  }\href@noop {} {\bibfield  {journal} {\bibinfo  {journal}
  {\url{https://www2.pt.tu-clausthal.de/paw/}}\ } (\bibinfo {year} {September
  28, 2016})}\BibitemShut {NoStop}%
\bibitem [{\citenamefont {Grimme}\ \emph {et~al.}(2010)\citenamefont {Grimme},
  \citenamefont {Antony}, \citenamefont {Ehrlich},\ and\ \citenamefont
  {Krieg}}]{Grimme2010}%
  \BibitemOpen
  \bibfield  {author} {\bibinfo {author} {\bibfnamefont {S.}~\bibnamefont
  {Grimme}}, \bibinfo {author} {\bibfnamefont {J.}~\bibnamefont {Antony}},
  \bibinfo {author} {\bibfnamefont {S.}~\bibnamefont {Ehrlich}},\ and\ \bibinfo
  {author} {\bibfnamefont {H.}~\bibnamefont {Krieg}},\ }\bibfield  {title}
  {\enquote {\bibinfo {title} {A consistent and accurate ab initio
  parametrization of density functional dispersion correction {(DFT-D)} for the
  94 elements {H-Pu}},}\ }\href@noop {} {\bibfield  {journal} {\bibinfo
  {journal} {J. Chem. Phys.}\ }\textbf {\bibinfo {volume} {132}},\ \bibinfo
  {pages} {154104} (\bibinfo {year} {2010})}\BibitemShut {NoStop}%
\bibitem [{\citenamefont {Grimme}, \citenamefont {Ehrlich},\ and\ \citenamefont
  {Goerigk}(2011)}]{Grimme2011}%
  \BibitemOpen
  \bibfield  {author} {\bibinfo {author} {\bibfnamefont {S.}~\bibnamefont
  {Grimme}}, \bibinfo {author} {\bibfnamefont {S.}~\bibnamefont {Ehrlich}},\
  and\ \bibinfo {author} {\bibfnamefont {L.}~\bibnamefont {Goerigk}},\
  }\bibfield  {title} {\enquote {\bibinfo {title} {Effect of the damping
  function in dispersion corrected density functional theory},}\ }\href@noop {}
  {\bibfield  {journal} {\bibinfo  {journal} {J. Comput. Chem.}\ }\textbf
  {\bibinfo {volume} {32}},\ \bibinfo {pages} {1456--1465} (\bibinfo {year}
  {2011})}\BibitemShut {NoStop}%
\bibitem [{\citenamefont {Behler}(2019)}]{RuNNer}%
  \BibitemOpen
  \bibfield  {author} {\bibinfo {author} {\bibfnamefont {J.}~\bibnamefont
  {Behler}},\ }\bibfield  {title} {\enquote {\bibinfo {title} {{RuNNer}},}\
  }\href@noop {} {\bibfield  {journal} {\bibinfo  {journal}
  {\url{http://gitlab.com/TheochemGoettingen/RuNNer}}\ } (\bibinfo {year}
  {August 22, 2019})}\BibitemShut {NoStop}%
\bibitem [{\citenamefont {Kalman}(1960)}]{Kalman1960}%
  \BibitemOpen
  \bibfield  {author} {\bibinfo {author} {\bibfnamefont {R.~E.}\ \bibnamefont
  {Kalman}},\ }\bibfield  {title} {\enquote {\bibinfo {title} {A new approach
  to linear filtering and prediction problems},}\ }\href@noop {} {\bibfield
  {journal} {\bibinfo  {journal} {J. Basic Eng.}\ }\textbf {\bibinfo {volume}
  {82}},\ \bibinfo {pages} {35--45} (\bibinfo {year} {1960})}\BibitemShut
  {NoStop}%
\bibitem [{\citenamefont {Blank}\ \emph {et~al.}(1995)\citenamefont {Blank},
  \citenamefont {Brown}, \citenamefont {Calhoun},\ and\ \citenamefont
  {Doren}}]{Blank1995}%
  \BibitemOpen
  \bibfield  {author} {\bibinfo {author} {\bibfnamefont {T.~B.}\ \bibnamefont
  {Blank}}, \bibinfo {author} {\bibfnamefont {S.~D.}\ \bibnamefont {Brown}},
  \bibinfo {author} {\bibfnamefont {A.~W.}\ \bibnamefont {Calhoun}},\ and\
  \bibinfo {author} {\bibfnamefont {D.~J.}\ \bibnamefont {Doren}},\ }\bibfield
  {title} {\enquote {\bibinfo {title} {Neural network models of potential
  energy surfaces},}\ }\href@noop {} {\bibfield  {journal} {\bibinfo  {journal}
  {J. Chem. Phys.}\ }\textbf {\bibinfo {volume} {103}},\ \bibinfo {pages}
  {4129--4137} (\bibinfo {year} {1995})}\BibitemShut {NoStop}%
\bibitem [{\citenamefont {Glorot}\ and\ \citenamefont
  {Bengio}(2010)}]{Glorot2010}%
  \BibitemOpen
  \bibfield  {author} {\bibinfo {author} {\bibfnamefont {X.}~\bibnamefont
  {Glorot}}\ and\ \bibinfo {author} {\bibfnamefont {Y.}~\bibnamefont
  {Bengio}},\ }\bibfield  {title} {\enquote {\bibinfo {title} {Understanding
  the difficulty of training deep feedforward neural networks},}\ }\href@noop
  {} {\bibfield  {journal} {\bibinfo  {journal} {J. Mach. Learn. Res.}\
  }\textbf {\bibinfo {volume} {9}},\ \bibinfo {pages} {249--256} (\bibinfo
  {year} {2010})}\BibitemShut {NoStop}%
\bibitem [{\citenamefont {Plimpton}(1995)}]{Plimpton1995}%
  \BibitemOpen
  \bibfield  {author} {\bibinfo {author} {\bibfnamefont {S.}~\bibnamefont
  {Plimpton}},\ }\bibfield  {title} {\enquote {\bibinfo {title} {Fast parallel
  algorithms for short-range molecular dynamics},}\ }\href@noop {} {\bibfield
  {journal} {\bibinfo  {journal} {J. Comput. Phys.}\ }\textbf {\bibinfo
  {volume} {117}},\ \bibinfo {pages} {1--19} (\bibinfo {year}
  {1995})}\BibitemShut {NoStop}%
\bibitem [{LAM(2019)}]{LAMMPS}%
  \BibitemOpen
  \bibfield  {title} {\enquote {\bibinfo {title} {{LAMMPS} -- {Large-scale}
  atomic/molecular massively parallel simulator},}\ }\href@noop {} {\bibfield
  {journal} {\bibinfo  {journal} {\url{http://lammps.sandia.gov}}\ } (\bibinfo
  {year} {August 7, 2019})}\BibitemShut {NoStop}%
\bibitem [{\citenamefont {Singraber}(2019)}]{n2p2}%
  \BibitemOpen
  \bibfield  {author} {\bibinfo {author} {\bibfnamefont {A.}~\bibnamefont
  {Singraber}},\ }\bibfield  {title} {\enquote {\bibinfo {title} {n2p2 -- {A}
  neural network potential package},}\ }\href@noop {} {\bibfield  {journal}
  {\bibinfo  {journal} {\url{https://github.com/CompPhysVienna/n2p2}}\ }
  (\bibinfo {year} {December 9, 2019})}\BibitemShut {NoStop}%
\bibitem [{\citenamefont {Nos{\'e}}(1984)}]{Nose1984}%
  \BibitemOpen
  \bibfield  {author} {\bibinfo {author} {\bibfnamefont {S.}~\bibnamefont
  {Nos{\'e}}},\ }\bibfield  {title} {\enquote {\bibinfo {title} {A molecular
  dynamics method for simulations in the canonical ensemble},}\ }\href@noop {}
  {\bibfield  {journal} {\bibinfo  {journal} {Mol. Phys.}\ }\textbf {\bibinfo
  {volume} {52}},\ \bibinfo {pages} {255--268} (\bibinfo {year}
  {1984})}\BibitemShut {NoStop}%
\bibitem [{\citenamefont {Hoover}(1985)}]{Hoover1985}%
  \BibitemOpen
  \bibfield  {author} {\bibinfo {author} {\bibfnamefont {W.~G.}\ \bibnamefont
  {Hoover}},\ }\bibfield  {title} {\enquote {\bibinfo {title} {Canonical
  dynamics: Equilibrium phase-space distributions},}\ }\href@noop {} {\bibfield
   {journal} {\bibinfo  {journal} {Phys. Rev. A}\ }\textbf {\bibinfo {volume}
  {31}},\ \bibinfo {pages} {1695--1697} (\bibinfo {year} {1985})}\BibitemShut
  {NoStop}%
\bibitem [{\citenamefont {Wales}\ and\ \citenamefont {Doye}(1997)}]{Wales1997}%
  \BibitemOpen
  \bibfield  {author} {\bibinfo {author} {\bibfnamefont {D.~J.}\ \bibnamefont
  {Wales}}\ and\ \bibinfo {author} {\bibfnamefont {J.~P.~K.}\ \bibnamefont
  {Doye}},\ }\bibfield  {title} {\enquote {\bibinfo {title} {Global
  optimization by basin-hopping and the lowest energy structures of
  {Lennard}-{Jones} clusters containing up to 110 atoms},}\ }\href@noop {}
  {\bibfield  {journal} {\bibinfo  {journal} {J. Phys. Chem. A}\ }\textbf
  {\bibinfo {volume} {101}},\ \bibinfo {pages} {5111--5116} (\bibinfo {year}
  {1997})}\BibitemShut {NoStop}%
\bibitem [{\citenamefont {Kanamori}(1960)}]{Kanamori1960}%
  \BibitemOpen
  \bibfield  {author} {\bibinfo {author} {\bibfnamefont {J.}~\bibnamefont
  {Kanamori}},\ }\bibfield  {title} {\enquote {\bibinfo {title} {Crystal
  distortion in magnetic compounds},}\ }\href@noop {} {\bibfield  {journal}
  {\bibinfo  {journal} {J. Appl. Phys.}\ }\textbf {\bibinfo {volume} {31}},\
  \bibinfo {pages} {S14--S23} (\bibinfo {year} {1960})}\BibitemShut {NoStop}%
\bibitem [{\citenamefont {Stukowski}(2010)}]{Ovito3.2.0}%
  \BibitemOpen
  \bibfield  {author} {\bibinfo {author} {\bibfnamefont {A.}~\bibnamefont
  {Stukowski}},\ }\bibfield  {title} {\enquote {\bibinfo {title} {Visualization
  and analysis of atomistic simulation data with {OVITO} -- the open
  visualization tool},}\ }\href@noop {} {\bibfield  {journal} {\bibinfo
  {journal} {Model. Simul. Mater. Sci. Eng.}\ }\textbf {\bibinfo {volume}
  {18}},\ \bibinfo {pages} {015012} (\bibinfo {year} {2010})}\BibitemShut
  {NoStop}%
\bibitem [{\citenamefont {Sun}\ \emph {et~al.}(2020)\citenamefont {Sun},
  \citenamefont {Liao}, \citenamefont {Wang}, \citenamefont {Chen},
  \citenamefont {Sun}, \citenamefont {Ong}, \citenamefont {Xi}, \citenamefont
  {Diao}, \citenamefont {Du}, \citenamefont {Wang}, \citenamefont {Breese},
  \citenamefont {Li}, \citenamefont {Zhang},\ and\ \citenamefont
  {Xu}}]{Sun2020}%
  \BibitemOpen
  \bibfield  {author} {\bibinfo {author} {\bibfnamefont {Y.}~\bibnamefont
  {Sun}}, \bibinfo {author} {\bibfnamefont {H.}~\bibnamefont {Liao}}, \bibinfo
  {author} {\bibfnamefont {J.}~\bibnamefont {Wang}}, \bibinfo {author}
  {\bibfnamefont {B.}~\bibnamefont {Chen}}, \bibinfo {author} {\bibfnamefont
  {S.}~\bibnamefont {Sun}}, \bibinfo {author} {\bibfnamefont {S.~J.~H.}\
  \bibnamefont {Ong}}, \bibinfo {author} {\bibfnamefont {S.}~\bibnamefont
  {Xi}}, \bibinfo {author} {\bibfnamefont {C.}~\bibnamefont {Diao}}, \bibinfo
  {author} {\bibfnamefont {Y.}~\bibnamefont {Du}}, \bibinfo {author}
  {\bibfnamefont {J.-O.}\ \bibnamefont {Wang}}, \bibinfo {author}
  {\bibfnamefont {M.~B.~H.}\ \bibnamefont {Breese}}, \bibinfo {author}
  {\bibfnamefont {S.}~\bibnamefont {Li}}, \bibinfo {author} {\bibfnamefont
  {H.}~\bibnamefont {Zhang}},\ and\ \bibinfo {author} {\bibfnamefont {Z.~J.}\
  \bibnamefont {Xu}},\ }\bibfield  {title} {\enquote {\bibinfo {title}
  {Covalency competition dominates the water oxidation structure-activity
  relationship on spinel oxides},}\ }\href@noop {} {\bibfield  {journal}
  {\bibinfo  {journal} {Nat. Catal.}\ }\textbf {\bibinfo {volume} {3}},\
  \bibinfo {pages} {554--563} (\bibinfo {year} {2020})}\BibitemShut {NoStop}%
\bibitem [{\citenamefont {Mills}\ and\ \citenamefont
  {J{\'o}nsson}(1994)}]{Mills1994}%
  \BibitemOpen
  \bibfield  {author} {\bibinfo {author} {\bibfnamefont {G.}~\bibnamefont
  {Mills}}\ and\ \bibinfo {author} {\bibfnamefont {H.}~\bibnamefont
  {J{\'o}nsson}},\ }\bibfield  {title} {\enquote {\bibinfo {title} {Quantum and
  thermal effects in {H$_2$} dissociative adsorption: Evaluation of free energy
  barriers in multidimensional quantum systems},}\ }\href@noop {} {\bibfield
  {journal} {\bibinfo  {journal} {Phys. Rev. Lett.}\ }\textbf {\bibinfo
  {volume} {72}},\ \bibinfo {pages} {1124--1127} (\bibinfo {year}
  {1994})}\BibitemShut {NoStop}%
\bibitem [{\citenamefont {J\'onsson}, \citenamefont {Mills},\ and\
  \citenamefont {Jacobsen}(1998)}]{Jonsson1998}%
  \BibitemOpen
  \bibfield  {author} {\bibinfo {author} {\bibfnamefont {H.}~\bibnamefont
  {J\'onsson}}, \bibinfo {author} {\bibfnamefont {G.}~\bibnamefont {Mills}},\
  and\ \bibinfo {author} {\bibfnamefont {K.~W.}\ \bibnamefont {Jacobsen}},\
  }\href@noop {} {\emph {\bibinfo {title} {Classical and Quantum Dynamics in
  Condensed Phase Simulations}}}\ (\bibinfo  {publisher} {World Scientific},\
  \bibinfo {address} {Singapore},\ \bibinfo {year} {1998})\BibitemShut
  {NoStop}%
\bibitem [{\citenamefont {Challa}, \citenamefont {Landau},\ and\ \citenamefont
  {Binder}(1986)}]{Challa1986}%
  \BibitemOpen
  \bibfield  {author} {\bibinfo {author} {\bibfnamefont {M.~S.~S.}\
  \bibnamefont {Challa}}, \bibinfo {author} {\bibfnamefont {D.~P.}\
  \bibnamefont {Landau}},\ and\ \bibinfo {author} {\bibfnamefont
  {K.}~\bibnamefont {Binder}},\ }\bibfield  {title} {\enquote {\bibinfo {title}
  {Finite-size effects at temperature-driven first-order transitions},}\
  }\href@noop {} {\bibfield  {journal} {\bibinfo  {journal} {Phys. Rev. B}\
  }\textbf {\bibinfo {volume} {34}},\ \bibinfo {pages} {1841--1852} (\bibinfo
  {year} {1986})}\BibitemShut {NoStop}%
\bibitem [{\citenamefont {Binder}(1987)}]{Binder1987}%
  \BibitemOpen
  \bibfield  {author} {\bibinfo {author} {\bibfnamefont {K.}~\bibnamefont
  {Binder}},\ }\bibfield  {title} {\enquote {\bibinfo {title} {Theory of
  first-order phase transitions},}\ }\href@noop {} {\bibfield  {journal}
  {\bibinfo  {journal} {Rep. Prog. Phys.}\ }\textbf {\bibinfo {volume} {50}},\
  \bibinfo {pages} {783--859} (\bibinfo {year} {1987})}\BibitemShut {NoStop}%
\bibitem [{\citenamefont {Ashcroft}\ and\ \citenamefont
  {Mermin}(1976)}]{Ashcroft1976}%
  \BibitemOpen
  \bibfield  {author} {\bibinfo {author} {\bibfnamefont {N.~W.}\ \bibnamefont
  {Ashcroft}}\ and\ \bibinfo {author} {\bibfnamefont {N.~D.}\ \bibnamefont
  {Mermin}},\ }\href@noop {} {\emph {\bibinfo {title} {Solid State Physics}}}\
  (\bibinfo  {publisher} {Saunders College},\ \bibinfo {address} {New York},\
  \bibinfo {year} {1976})\BibitemShut {NoStop}%
\bibitem [{\citenamefont {Bisquert}(2008)}]{Bisquert2008}%
  \BibitemOpen
  \bibfield  {author} {\bibinfo {author} {\bibfnamefont {J.}~\bibnamefont
  {Bisquert}},\ }\bibfield  {title} {\enquote {\bibinfo {title} {Interpretation
  of electron diffusion coefficient in organic and inorganic semiconductors
  with broad distributions of states},}\ }\href@noop {} {\bibfield  {journal}
  {\bibinfo  {journal} {Phys. Chem. Chem. Phys.}\ }\textbf {\bibinfo {volume}
  {10}},\ \bibinfo {pages} {3175–3194} (\bibinfo {year} {2008})}\BibitemShut
  {NoStop}%
\bibitem [{\citenamefont {Bisquert}(2004)}]{Bisquert2004}%
  \BibitemOpen
  \bibfield  {author} {\bibinfo {author} {\bibfnamefont {J.}~\bibnamefont
  {Bisquert}},\ }\bibfield  {title} {\enquote {\bibinfo {title} {Chemical
  diffusion coefficient of electrons in nanostructured semiconductor electrodes
  and dye-sensitized solar cells},}\ }\href@noop {} {\bibfield  {journal}
  {\bibinfo  {journal} {J. Phys. Chem. B}\ }\textbf {\bibinfo {volume} {108}},\
  \bibinfo {pages} {2323--2332} (\bibinfo {year} {2004})}\BibitemShut {NoStop}%
\bibitem [{\citenamefont {Mehrer}(2007)}]{Mehrer2007}%
  \BibitemOpen
  \bibfield  {author} {\bibinfo {author} {\bibfnamefont {H.}~\bibnamefont
  {Mehrer}},\ }\href@noop {} {\emph {\bibinfo {title} {Diffusion in Solids}}}\
  (\bibinfo  {publisher} {Springer},\ \bibinfo {address} {Berlin},\ \bibinfo
  {year} {2007})\BibitemShut {NoStop}%
\bibitem [{\citenamefont {Phillips}(1989)}]{Phillips1989}%
  \BibitemOpen
  \bibfield  {author} {\bibinfo {author} {\bibfnamefont {J.~C.}\ \bibnamefont
  {Phillips}},\ }\href@noop {} {\emph {\bibinfo {title} {Physics of
  High-{$T_\mathrm{c}$} Superconductors}}}\ (\bibinfo  {publisher} {Academic
  Press},\ \bibinfo {address} {San Diego},\ \bibinfo {year} {1989})\BibitemShut
  {NoStop}%
\end{thebibliography}%

\end{document}